\documentclass[aps,prb,reprint,superscriptaddress,floatfix]{revtex4-2}
\usepackage[normalem]{ulem}

\usepackage{xcolor}
\usepackage{graphicx}
\usepackage{amsmath}
\usepackage{physics}
\usepackage[colorlinks=true,allcolors=blue]{hyperref}

\begin{document}

\title{Fluctuating intertwined stripes in the strange metal regime of the Hubbard model}

\author{Edwin W. Huang}
\thanks{These authors contributed equally.}
\affiliation{Department of Physics and Institute of Condensed Matter Theory,
University of Illinois at Urbana-Champaign, Urbana, IL 61801, USA.
}

\author{Tianyi Liu}
\thanks{These authors contributed equally.}
\affiliation{Stanford Institute for Materials and Energy Sciences,
SLAC National Accelerator Laboratory, 2575 Sand Hill Road, Menlo Park, CA 94025, USA
}
\affiliation{Department of Chemistry, Stanford University, Stanford, CA
94305, USA}
\author{Wen O. Wang}
\affiliation{Stanford Institute for Materials and Energy Sciences,
SLAC National Accelerator Laboratory, 2575 Sand Hill Road, Menlo Park, CA 94025, USA
}

\affiliation{Department of Applied Physics, Stanford University, Stanford, CA
94305, USA}
\author{Hong-Chen Jiang}
\affiliation{Stanford Institute for Materials and Energy Sciences,
SLAC National Accelerator Laboratory, 2575 Sand Hill Road, Menlo Park, CA 94025, USA
}

\author{Peizhi Mai}
\affiliation{Department of Physics and Institute of Condensed Matter Theory,
University of Illinois at Urbana-Champaign, Urbana, IL 61801, USA.
}

\author{Thomas A. Maier}
\affiliation{Computational Sciences and Engineering Division, Oak Ridge National Laboratory, Oak Ridge, TN, 37831-6494, USA\looseness=-1}

\author{Steven Johnston}
\affiliation{Department of Physics and Astronomy, The University of Tennessee, Knoxville, Tennessee 37996, USA}
\affiliation{Institute of Advanced Materials and Manufacturing, The University of Tennessee, Knoxville, TN 37996, USA\looseness=-1} 

\author{Brian Moritz}
\affiliation{Stanford Institute for Materials and Energy Sciences,
SLAC National Accelerator Laboratory, 2575 Sand Hill Road, Menlo Park, CA 94025, USA
}

\author{Thomas P. Devereaux}
\affiliation{Stanford Institute for Materials and Energy Sciences,
SLAC National Accelerator Laboratory, 2575 Sand Hill Road, Menlo Park, CA 94025, USA
}
\affiliation{Department of Materials Science and Engineering, Stanford University, Stanford, CA 94305, USA}

\date{\today}

\maketitle

\textbf{Strongly correlated electron systems host a variety of poorly understood correlations in their high temperature normal state. Unlike ordered phases defined by order parameters, these normal state phases are often defined through unconventional properties such as strange metallic transport or spectroscopic pseudogaps. Characterizing the microscopic correlations in the normal state is necessary to elucidate mechanisms that lead to these properties and their connection to ground state orders. Here we establish the presence of intertwined charge and spin stripes in the strange metal normal state of the Hubbard model using determinant quantum Monte Carlo calculations. The charge and spin density waves constituting the stripes are fluctuating and short-ranged, yet they obey a mutual commensurability relation and remain microscopically interlocked, as evidenced through measurements of three-point spin-spin-hole correlation functions. Our findings demonstrate the ability of many-body numerical simulations to unravel the microscopic correlations that define quantum states of matter.}

The concept of intertwined orders is commonly used to characterize states within the pseudogap regime of the cuprate phase diagram~\cite{fradkin2012,fradkin2015}. A well-known example is that of stripe order, unidirectional spin and charge density waves that are most prominent at low temperatures around $p=1/8$ hole doping~\cite{kivelson2003,huang2017,zheng2017,Huang2018}. In La-based cuprates~\cite{tranquada1995,tranquada2021} and in simulations of the Hubbard model~\cite{zheng2017}, spin and charge stripes are interlocked. Regions of high hole concentration are aligned with antiferromagnetic phase reversals. Stripe order is well known to interact closely with superconductivity, as evidenced by $1/8$-anomalies in cuprate experiments~\cite{feng2006} and by nearly-degenerate ground state energies in Hubbard model calculations~\cite{Corboz2014,zheng2017}. The close interplay of spin and charge orders and their competition with superconductivity are believed to be hallmarks of the pseudogap regime.

The majority of recent progress in solving the Hubbard model has targeted ground state properties~\cite{arovas2022,qin2022,jiang2019,qin2020,Chung2020,jiang2020,sorella2021}. Studies at finite temperature have found fluctuating spin and charge stripes~\cite{Huang2018,wietek2021,mai2021}, but their interplay, doping dependence, and placement in the broader phase diagram have not been explored thoroughly. Our calculations of the Hubbard model demonstrate interlocked spin and charge stripes at temperatures above the onset of the pseudogap, in the strange metal regime characterized by $T$-linear resistivity~\cite{brown2019,huang2019}. The wide range of doping where we find stripes corroborate a growing number of experimental studies finding charge stripes in optimally doped and overdoped cuprates~\cite{arpaia2019,miao2021,lin2020,peng2018,lee2020,ma2021,tam2022,lee2021,kawasaki2021}.

\begin{figure}
    \includegraphics{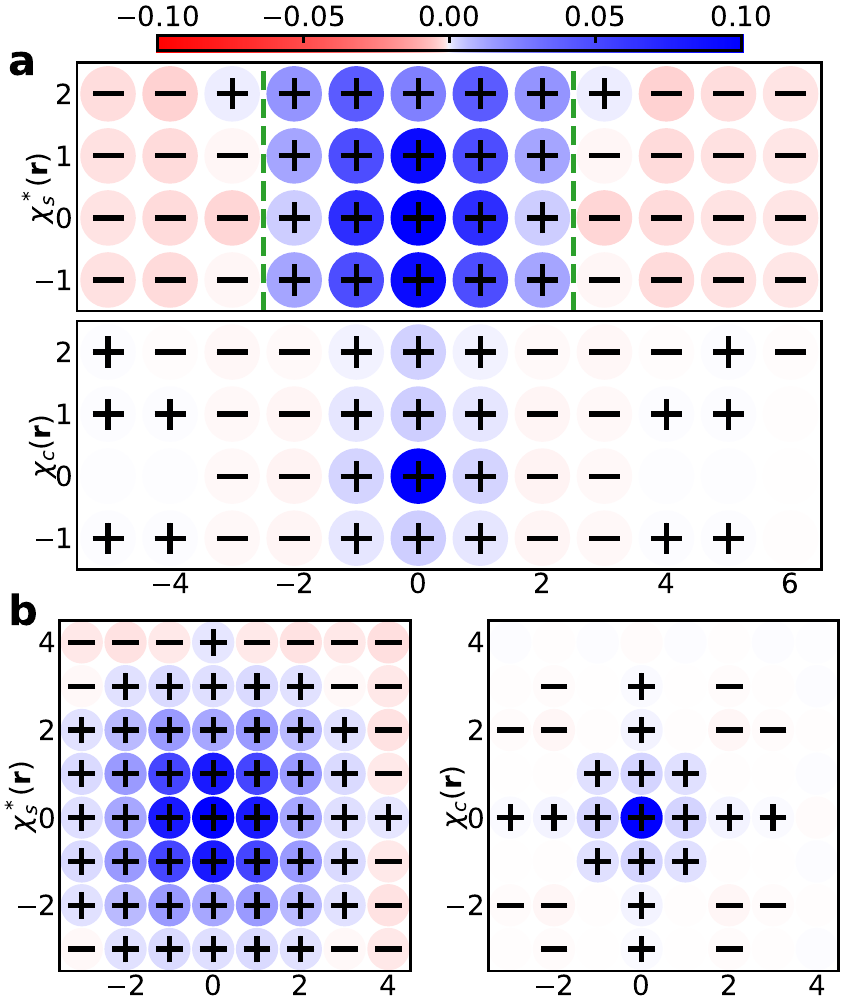}
    \caption{\textbf{Stripes in spin and charge susceptibilities}. \textbf{a,b,} Spin susceptibility $\chi_s(\mathbf{r})$ and charge susceptibility $\chi_c(\mathbf{r})$ at zero frequency in the Hubbard model. The spin susceptibility is plotted with a staggering factor for clarity ($\chi^*_s(\mathbf{r}) = \chi_s(\mathbf{r}) \times (-1)^{r_x + r_y}$). Parameters are $U/t = 6, t^\prime/t = -0.25, T/t \approx 0.22, p=0.125$. Cluster size is \textbf{a} $12\times4$ and \textbf{b} $8\times8$. + and - signs indicate correlations that are nonzero by at least two standard errors. Green dashed lines in \textbf{a} denote the antiphase domain walls of spin stripes. The diamond patterns of modulation in \textbf{b} indicate a superposition of stripes along $x$ and $y$ directions.}
    \label{fig:1}
\end{figure}

Our results are based on unbiased determinant quantum Monte Carlo (DQMC) simulations~\cite{white1989,bss1981} conducted with very large sample sizes. Typical simulations involve $\sim 10^{10}$ measurements, allowing for small stochastic errors ($\sim 10^{-6}$) despite the presence of a fermion sign problem. The principal observables we compute to investigate stripes in the Hubbard model are the charge and spin susceptibilities at zero frequency, defined as 
\begin{align}
\chi_c(\mathbf{r}) &= \int_0^\beta \dd{\tau} \ev{n_{\mathbf{r}}(\tau) n_{\mathbf{0}}} - \ev{n_{\mathbf{r}}}\ev{n_{\mathbf{0}}} \label{defchic_} \\
\chi_s(\mathbf{r}) &= \int_0^\beta \dd{\tau} \ev{m^z_{\mathbf{r}}(\tau) m^z_{\mathbf{0}}} \label{defchis_},
\end{align}
where $n_{\mathbf{r}} = n_{\mathbf{r} \uparrow} + n_{\mathbf{r} \downarrow}$ and $m^z_{\mathbf{r}} = \frac{1}{2}\qty(n_{\mathbf{r} \uparrow} - n_{\mathbf{r} \downarrow})$ are the charge and spin densities on site $\mathbf{r}$. These quantities can be computed directly with DQMC, without the need for analytic continuation, so that our results are numerically exact. 

Figure \ref{fig:1}{\bf a} displays the spin and charge susceptibilities as functions of $\mathbf{r}$ for a $12 \times 4$ rectangular cluster with periodic boundary conditions at doping $p=1/8$. The spin susceptibility is plotted with a staggering factor ($\chi^*_s(\mathbf{r}) = \chi_s(\mathbf{r}) \times (-1)^{r_x + r_y}$) to highlight deviations from commensurate antiferromagnetism. In both the spin and charge susceptibilities, periodic modulations are visible along the long direction of the cluster, indicating the presence of short-ranged fluctuating stripes. The period of the charge modulation is approximately half of that of the spin modulation, consistent with a stripe pattern where antiphase domain walls in the spin density coincide with regions of increased hole density~\cite{zaanen1989,zaanen2001}. The pattern of modulation in the spin susceptibility is identical to that in the equal-time $(\tau = 0)$ spin correlation function (Fig.~\ref{fig:chi_S_12x4}, \ref{fig:chi_S_16x4}), analyzed previously in Ref.~\cite{Huang2018}. By contrast, the stripe modulations in the charge susceptibility are not visible in the equal-time charge correlation function~\cite{mai2021}, at the temperatures attainable in our simulations. This distinction is related to the fact that high-energy incoherent excitations contaminate the equal-time correlation function more than the static susceptibility, as emphasized in Ref.~\cite{kivelson2003}.

\begin{figure*}
    \includegraphics{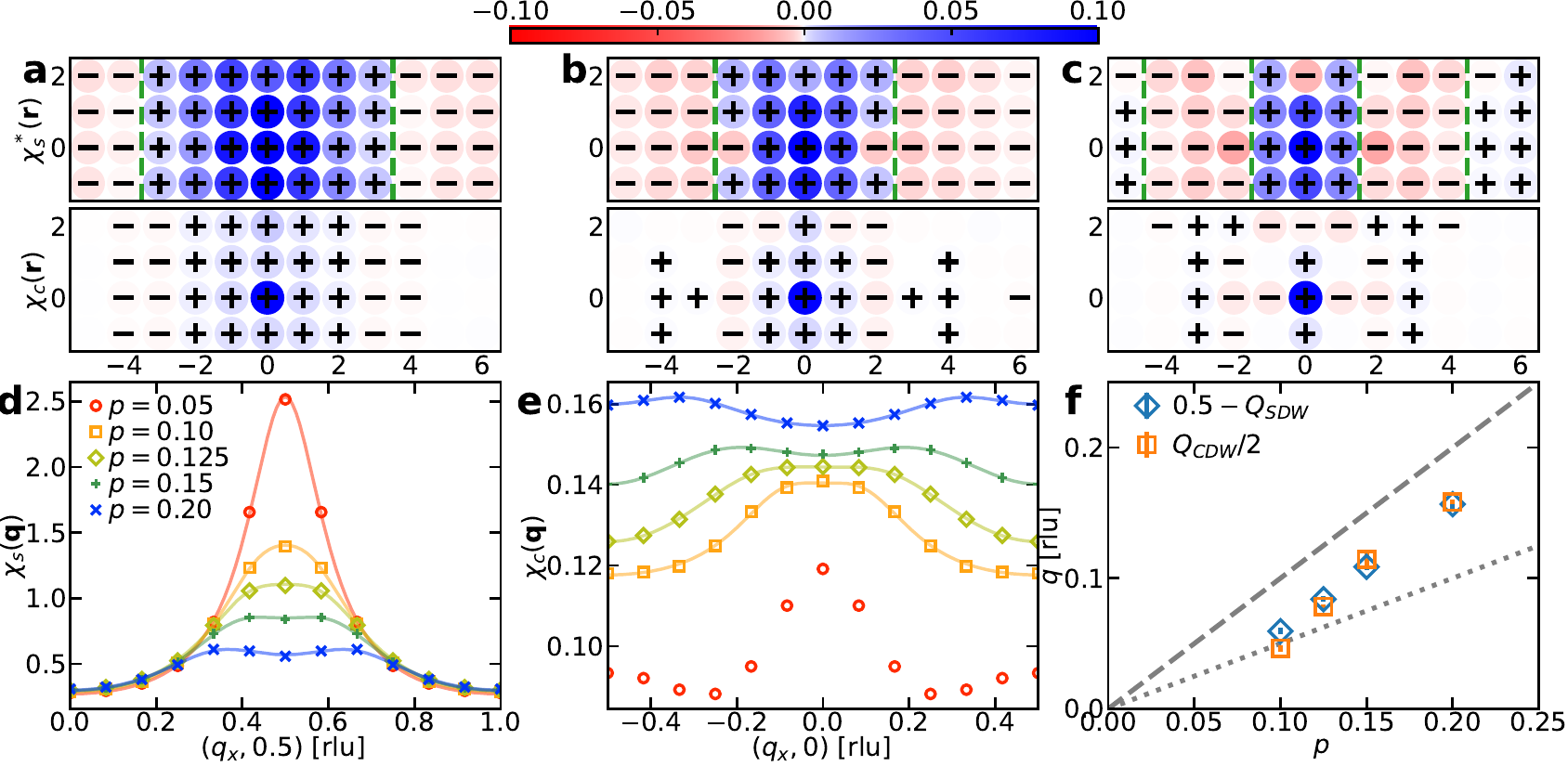}
    \caption{\textbf{Doping dependence of spin and charge stripes.} \textbf{a, b, c,} Staggered spin susceptibilities and charge susceptibilities for dopings $0.1$, $0.15$, and $0.2$, respectively, on a $12\times4$ cluster with parameters $U/t = 6, t^\prime/t = -0.25, T/t \approx 0.22$. \textbf{d, e,} momentum-space susceptibilities for spin and charge respectively for various doping concentrations. Solid lines indicate fits to periodic Lorentzian functions (See Supplement \ref{sec:fitting}). No stripes are present at $p=0.05$, and the Lorentzian fit to $\chi_c(\mathbf{q})$ is poor and not shown. \textbf{f,} Spin and charge incommensurabilities as a function of doping, obtained from fits to the momentum-space susceptibilities. Dashed and dotted lines indicate $q=p$ and $q=p/2$, corresponding to the spin incommensurabilities of half-filled and filled stripes, respectively.}
    \label{fig:2}
\end{figure*}

We have checked that the finite size cluster does not have a notable impact on the properties of the stripe pattern (Figs.~\ref{fig:chi_S_12x4}-\ref{fig:Nx_dep}). We focus on $12\times4$ cluster results in Fig.~\ref{fig:1}{\bf a} and Fig.~\ref{fig:2} as the larger average fermion sign associated with smaller cluster size enables us to more clearly resolve modulations in the charge susceptibility. We consider an $8\times8$ cluster in Fig.~\ref{fig:1}{\bf b}. Here, modulations are again visible in both the spin and charge susceptibilties, with negative regions along the diagonal directions. This pattern is precisely expected from a superposition of horizontal and vertical stripes. Our analysis indicates that the stripe modulations seen for the $12\times4$ cluster are not artifacts of limited system size.

The doping dependence of the spin and charge susceptibilities is shown in Fig.~\ref{fig:2}. In Fig.~\ref{fig:2}{\bf a-c}, we plot the susceptibilities for hole doping concentrations of $p=0.1, 0.15,$ and $0.2$. Modulations are present, indicating fluctuating spin and charge stripes for all three doping levels. The period of the modulation decreases with increased hole doping. This is also clearly reflected in momentum-space susceptibilities. In Fig.~\ref{fig:2}{\bf d}, the spin susceptibility splits from a single peak at $(\pi, \pi)$ (i.e.\ $(0.5, 0.5)$ in reciprocal lattice units) to two incommensurate peaks with increased hole doping. The data are well fit with periodic Lorentzian functions (Supplement \ref{sec:fitting}). Similarly, the charge susceptibility splits away from $\mathbf{q} = (0, 0)$ as hole doping increases and rises uniformly owing to the increased metallicity of the doped system. For hole doping $0.1 \leq p \leq 0.2$ we obtain excellent fits to $\chi_c(\mathbf{q})$ with periodic Lorentzian functions plus a constant background. In Fig.~\ref{fig:nnzzqw0s} \textbf{b}, we check that the charge susceptibilities are indeed peaked close to $(0,0)$ or $(\pi, 0)$, rather than near $(\pi, \pi)$ as in the non-interacting model. This indicates that the fluctuating stripes we observe are unrelated to Fermi surface effects such as nesting.

From our fits to $\chi_s(\mathbf{q})$ and $\chi_c(\mathbf{q})$, we extract the spin and charge incommensurabilities, defined as the separation of the incommensurate peaks from the commensurate wavevectors ($(0,0)$ for charge, and $(\pi,\pi)$ for spin). Fig.~\ref{fig:2}{\bf f} plots the spin and charge incommensurabilities against doping. The spin incommensurability is very close to half the charge incommensurability through the range of doping $0.1 \leq p \leq 0.2$, indicating that the stripes are mutually commensurate. Both increase monotonically with hole doping, with a stripe filling in between half-filled (dashed line) and fully-filled (dotted line).

\begin{figure}
    \includegraphics{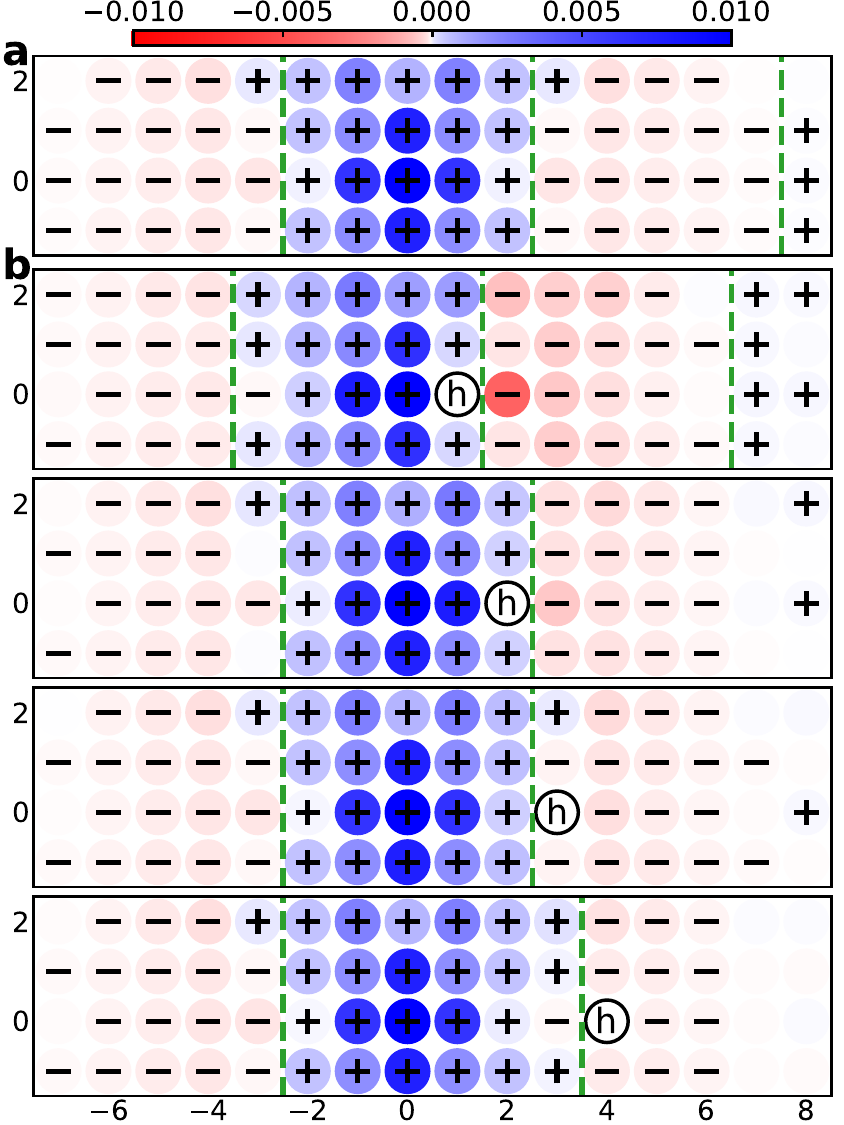}
    \caption{\textbf{Holes pin antiphase domain walls}. \textbf{a,} Uncorrelated value $\ev{m^z_\mathbf{r} m^z_\mathbf{0}}\ev{h_{\mathbf{r}'}}$ demonstrating expectations if spin and charge were decoupled. As in Figs.~\ref{fig:1} and \ref{fig:2}, a staggering factor $(-1)^{r_x + r_y}$ is included for clarity. \textbf{b,} Full spin-spin-hole correlation function $\ev{m^z_\mathbf{r} m^z_\mathbf{0} h_{\mathbf{r}'}}$ indicating spin correlations in the presence of a hole. The location of the hole on $\mathbf{r}'$ is indicated by the letter `h'. The antiphase domain walls (green dashed lines) move as $\mathbf{r}'$ is varied. Parameters are $U/t=6$, $t'/t=-0.25$, $p = 0.125$, $T/t=0.22$.}
    \label{fig:3}
\end{figure}

The mutual commensurability of the spin and charge stripes strongly suggests, but does not prove, that doped holes reside near antiphase domain walls. It is known that modifying the chemical potential on a column can pin the location of antiphase domain walls~\cite{mondaini2012}, and conversely that including a staggered magnetic field on a column can induce a static charge stripe modulation~\cite{mai2021}. However, whether spin and charge stripes are pinned to each other while still fluctuating is unknown. To resolve this question and probe the relation between fluctuating spin and charge stripes, we consider the $3$-point spin-spin-hole correlation function
\begin{equation}
    \ev{m^z_\mathbf{r} m^z_\mathbf{0} h_{\mathbf{r}'}} \label{def_ssh}
\end{equation}
where $h_{\mathbf{r}'} = c_{\mathbf{r}'\uparrow} c_{\mathbf{r}'\uparrow}^\dagger c_{\mathbf{r}'\downarrow} c_{\mathbf{r}'\downarrow}^\dagger$ ensures the presence of a hole on site $\mathbf{r}'$. In Fig.~\ref{fig:3}{\bf a} we first plot $\ev{m^z_\mathbf{r} m^z_\mathbf{0}}\ev{h_{\mathbf{r}'}}$ to demonstrate how the $3$-point correlation function would appear if spin and charge were entirely decoupled. By translation symmetry, $\ev{h_{\mathbf{r}'}}$ is a constant and Fig.~\ref{fig:3}{\bf a} thus simply shows the spin correlation function. Figure \ref{fig:3}{\bf b} shows the full $3$-point correlation function, with the axes and letter ``h'' indicating the coordinates of $\mathbf{r}$ and $\mathbf{r}'$ respectively. In these $3$-point correlation functions, it is clear that while the periodicity of spin stripes is unaffected, there is a strong tendency for the antiphase domain walls to lie adjacent to the hole. This establishes definitively that although both the spin and charge stripes seen in Figs.~\ref{fig:1} and \ref{fig:2} are short-ranged and fluctuating, they remain microscopically interlocked. This close interplay between fluctuating spin and charge stripes in our finite temperature calculations indicates that the notion of intertwined orders is not unique to the pseudogap regime of the phase diagram.

\begin{figure*}
    \includegraphics[scale=0.6]{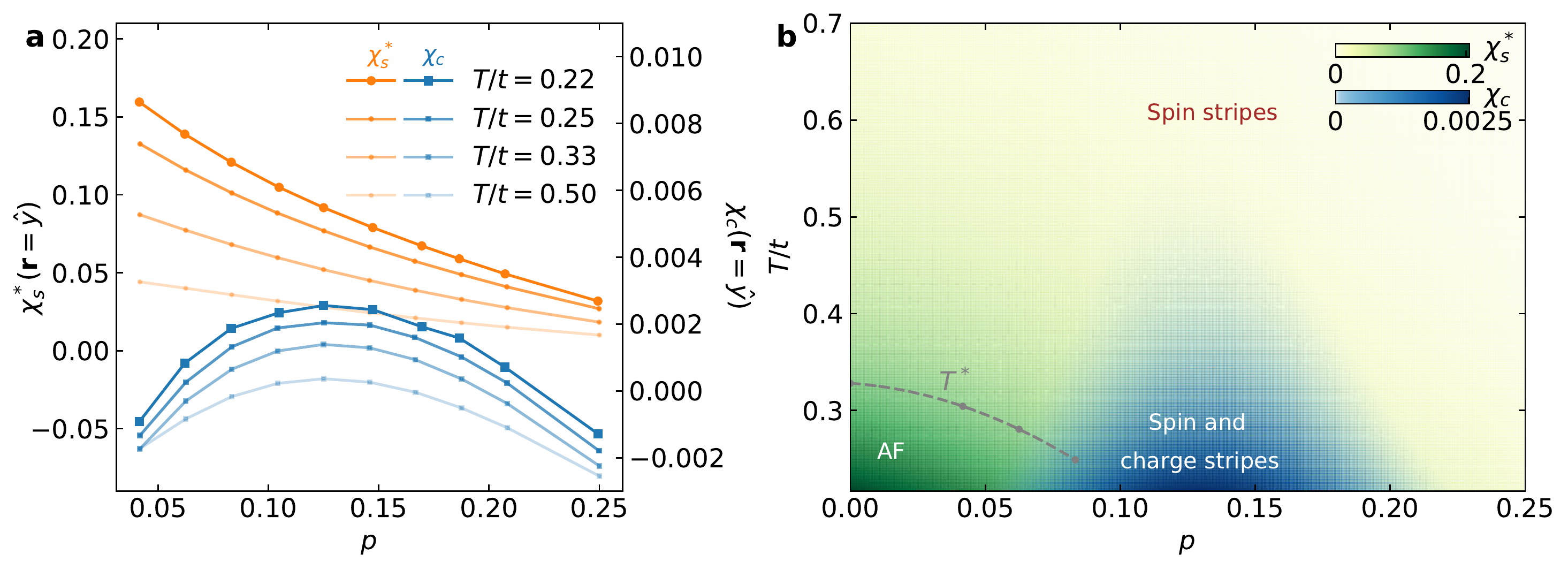}
    \caption{\textbf{Doping and temperature dependence of stripe intensity.} \textbf{a}, Staggered spin [$\chi^*_s(\boldsymbol{r}=\hat{y})$] and charge [$\chi_c(\boldsymbol{r}=\hat{y})$] susceptibilities at the y-neighbor, as a function of doping for different temperatures. Lines are guides to the eye. \textbf{b}, Schematic temperature-doping phase diagram drawn using data from \textbf{a}. The yellow-green background corresponds to $\chi^*_s(\hat{y})$, indicating local antiferromagnetic correlations throughout the temperature doping ranges. Darker colors indicate larger magnitude. The blue region indicates positive $\chi_c(\hat{y})$, and the color intensity corresponds to the value of the susceptibility. Cubic spline interpolation is applied to data points in the temperature-doping grid to obtain smoothly varying color intensity. Dashed grey line indicates the approximate crossover temperatures of the pseudogap regime, $T^*$, estimated by locating the temperatures that maximize the spin (Pauli) susceptibilities or Knight shift for various dopings (Fig.~\ref{fig:pauli}).}
    \label{fig:phase_diagram}
\end{figure*}

In the data presented thus far, we have focused on results at a temperature $T/t \approx 0.22$, near the lowest accessible in our simulations due to the fermion sign problem. In Fig.~\ref{fig:phase_diagram}, we discuss the evolution of the strength of the fluctuating stripes with temperature and doping. We consider the value of the spin and charge susceptibilities at the nearest vertical neighbor, $\chi_s(\mathbf{r} = \hat{y})$ and $\chi_c(\mathbf{r} = \hat{y})$, as simple estimates of the magnitude of the fluctuating stripes (See Supplement \ref{sec:stripe_strength}).

The doping and temperature dependence of $\chi_{c,s}(\mathbf{r} = \hat{y})$ is plotted in Fig.~\ref{fig:phase_diagram}{\bf a}. For doping concentrations or temperatures where $\chi_c(\mathbf{r} = \hat{y}) < 0$, the patterns in the charge susceptibility do not resemble stripes (Fig.~\ref{fig:Tdep}). We observe that the spin stripes weaken monotonically with increasing doping and increasing temperature, but the charge stripes display a non-monotonic doping dependence with a maximum at $p=1/8$. This peak is highly reminiscent of $1/8$-anomalies in cuprate superconductors, where charge stripes have also been observed to have maximal strength at $p=1/8$, with a concomitant suppression of superconductivity. The broad peak at $p=1/8$ is also reminiscent of a similar peak seen in a $\mathbf{q}=0$ nematic susceptibility reported previously for the Hubbard model~\cite{liu2021}.

Both spin and charge stripes grow in intensity as temperature decreases. While there is no sharp definition for the onset temperature of fluctuating stripes, given their short-ranged nature, we generally find that modulations indicative of charge stripes onset at lower temperatures than spin stripes. Our findings are summarized in a temperature-doping ``phase diagram" in Fig.~\ref{fig:phase_diagram}{\bf b}, where the color intensity corresponds to the magnitude of $\chi_s(\mathbf{r}=\hat{y})$ in the yellow/green background, on top of which a blue region around $p=1/8$ corresponding to $\chi_c(\mathbf{r}=\hat{y})$ is overlaid. In general, we find that incommensurate spin correlations indicative of spin stripes become visible below roughly $T/t \approx 0.6$. Charge stripes become visible at lower temperatures, for instance $T/t \approx 0.5$ at $p=1/8$.

We emphasize that the clear and robust signatures of interlocked spin and charge stripes occur at temperatures well above the onset of the pseudogap. The pseudogap crossover temperature $T^*$ is estimated by the peak in the Knight shift $\chi_s(\mathbf{q}=0, \omega=0)$ as a function of temperature (See Supplement~\ref{sec:PG}). $T^*$ for different doping is plotted in Fig.~\ref{fig:phase_diagram}{\bf b}. As $T^*$ decreases with increased hole doping, we cannot explore the behavior of fluctuating stripes below the pseudogap onset temperature within unbiased DQMC simulations. Nevertheless we find strong signatures of fluctuating spin and charge stripes over a significant range of hole doping, thus demonstrating that the pseudogap is not a prerequisite for intertwined orders. In fact, the temperatures and doping levels at which our simulations are conducted lie in the strange metal regime of the phase diagram, as supported by previous DQMC calculations of the Hubbard model finding large, $T$-linear resistivity~\cite{huang2019}. Our findings motivate further studies and analysis of theories connecting fluctuating stripes to strange metallic transport ~\cite{luca2017scipost,luca2017prb,andrade2018,caprara2022}.

The presence of stripes in the strange metal regime is further substantiated by a number of recent X-ray scattering experiments finding scattering from fluctuating charge density waves in optimally and over-doped cuprate compounds at temperatures approaching room temperature~\cite{arpaia2019,miao2021,lin2020,peng2018,lee2020,ma2021,tam2022,lee2021}. A recent detailed analysis of Eu-LSCO~\cite{lee2021} showed a nearly temperature-independent integrated intensity of the charge scattering peak over range of doping $0.1 \leq p \leq 0.2$, indicating that the amplitude of the charge order onsets above experimentally accessible temperatures and that stripes persist well into the strange metal. Interestingly, the same study suggests decoupling of spin and charge stripes at elevated temperatures, in contrast to the results for the Hubbard model presented here. While the Hubbard model is clearly insufficient for a realistic description of the cuprates' electronic structure, the concordance of our numerical results and recent experimental works highlights the importance of fluctuating stripes over a larger region of the phase diagram than previously considered. Their existence over wide ranges of doping and temperatures is evidence of their relevance to all electronic properties of cuprates. Our findings call for further investigations of intertwined order in other strongly correlated materials and strange metals.

\begin{acknowledgments}
The work at Stanford and SLAC (T.L., W.O.W., H.C.J., B.M., and T.P.D.) was supported by the US Department of Energy, Office of Basic Energy Sciences, Materials Sciences and Engineering Division, under Contract No. DE-AC02-76SF00515. E.W.H. was supported by the Gordon and Betty Moore Foundation EPiQS Initiative through the grants GBMF 4305 and GBMF 8691. PM acknowledges support from Center for Emergent Superconductivity, a DOE Energy Frontier Research Center, Grant No. DE-AC0298CH1088. Work by S.J. and T.A.M. was supported by the U.S. Department of Energy, Office of Science, Office of Basic Energy Sciences, under Award Number DE-SC0022311. 
Computational work was performed on the Sherlock cluster at Stanford University and on resources of the National Energy Research Scientific Computing Center, supported by the U.S. Department of Energy under contract DE-AC02-05CH11231.
\end{acknowledgments}

\textbf{Competing Interests:} The authors declare no competing financial or non-financial interests.

\textbf{Data Availability:} The data that support the findings of this study are available from the corresponding author upon reasonable request.

\textbf{Author contributions:} E.W.H., T.L., and W.O.W. performed DQMC simulations. E.W.H and T.L. analyzed the data. T.P.D. and B.M. supervised the project. All authors discussed the results and participated in writing the manuscript.


\begin{thebibliography}{41}%
\makeatletter
\providecommand \@ifxundefined [1]{%
 \@ifx{#1\undefined}
}%
\providecommand \@ifnum [1]{%
 \ifnum #1\expandafter \@firstoftwo
 \else \expandafter \@secondoftwo
 \fi
}%
\providecommand \@ifx [1]{%
 \ifx #1\expandafter \@firstoftwo
 \else \expandafter \@secondoftwo
 \fi
}%
\providecommand \natexlab [1]{#1}%
\providecommand \enquote  [1]{``#1''}%
\providecommand \bibnamefont  [1]{#1}%
\providecommand \bibfnamefont [1]{#1}%
\providecommand \citenamefont [1]{#1}%
\providecommand \href@noop [0]{\@secondoftwo}%
\providecommand \href [0]{\begingroup \@sanitize@url \@href}%
\providecommand \@href[1]{\@@startlink{#1}\@@href}%
\providecommand \@@href[1]{\endgroup#1\@@endlink}%
\providecommand \@sanitize@url [0]{\catcode `\\12\catcode `\$12\catcode
  `\&12\catcode `\#12\catcode `\^12\catcode `\_12\catcode `\%12\relax}%
\providecommand \@@startlink[1]{}%
\providecommand \@@endlink[0]{}%
\providecommand \url  [0]{\begingroup\@sanitize@url \@url }%
\providecommand \@url [1]{\endgroup\@href {#1}{\urlprefix }}%
\providecommand \urlprefix  [0]{URL }%
\providecommand \Eprint [0]{\href }%
\providecommand \doibase [0]{https://doi.org/}%
\providecommand \selectlanguage [0]{\@gobble}%
\providecommand \bibinfo  [0]{\@secondoftwo}%
\providecommand \bibfield  [0]{\@secondoftwo}%
\providecommand \translation [1]{[#1]}%
\providecommand \BibitemOpen [0]{}%
\providecommand \bibitemStop [0]{}%
\providecommand \bibitemNoStop [0]{.\EOS\space}%
\providecommand \EOS [0]{\spacefactor3000\relax}%
\providecommand \BibitemShut  [1]{\csname bibitem#1\endcsname}%
\let\auto@bib@innerbib\@empty
\bibitem [{\citenamefont {Fradkin}\ and\ \citenamefont
  {Kivelson}(2012)}]{fradkin2012}%
  \BibitemOpen
  \bibfield  {author} {\bibinfo {author} {\bibfnamefont {E.}~\bibnamefont
  {Fradkin}}\ and\ \bibinfo {author} {\bibfnamefont {S.~A.}\ \bibnamefont
  {Kivelson}},\ }\href {https://doi.org/10.1038/nphys2498} {\bibfield
  {journal} {\bibinfo  {journal} {Nature Physics}\ }\textbf {\bibinfo {volume}
  {8}},\ \bibinfo {pages} {864} (\bibinfo {year} {2012})}\BibitemShut {NoStop}%
\bibitem [{\citenamefont {Fradkin}\ \emph {et~al.}(2015)\citenamefont
  {Fradkin}, \citenamefont {Kivelson},\ and\ \citenamefont
  {Tranquada}}]{fradkin2015}%
  \BibitemOpen
  \bibfield  {author} {\bibinfo {author} {\bibfnamefont {E.}~\bibnamefont
  {Fradkin}}, \bibinfo {author} {\bibfnamefont {S.~A.}\ \bibnamefont
  {Kivelson}},\ and\ \bibinfo {author} {\bibfnamefont {J.~M.}\ \bibnamefont
  {Tranquada}},\ }\href {https://doi.org/10.1103/RevModPhys.87.457} {\bibfield
  {journal} {\bibinfo  {journal} {Rev. Mod. Phys.}\ }\textbf {\bibinfo {volume}
  {87}},\ \bibinfo {pages} {457} (\bibinfo {year} {2015})}\BibitemShut
  {NoStop}%
\bibitem [{\citenamefont {Kivelson}\ \emph {et~al.}(2003)\citenamefont
  {Kivelson}, \citenamefont {Bindloss}, \citenamefont {Fradkin}, \citenamefont
  {Oganesyan}, \citenamefont {Tranquada}, \citenamefont {Kapitulnik},\ and\
  \citenamefont {Howald}}]{kivelson2003}%
  \BibitemOpen
  \bibfield  {author} {\bibinfo {author} {\bibfnamefont {S.~A.}\ \bibnamefont
  {Kivelson}}, \bibinfo {author} {\bibfnamefont {I.~P.}\ \bibnamefont
  {Bindloss}}, \bibinfo {author} {\bibfnamefont {E.}~\bibnamefont {Fradkin}},
  \bibinfo {author} {\bibfnamefont {V.}~\bibnamefont {Oganesyan}}, \bibinfo
  {author} {\bibfnamefont {J.~M.}\ \bibnamefont {Tranquada}}, \bibinfo {author}
  {\bibfnamefont {A.}~\bibnamefont {Kapitulnik}},\ and\ \bibinfo {author}
  {\bibfnamefont {C.}~\bibnamefont {Howald}},\ }\href
  {https://doi.org/10.1103/RevModPhys.75.1201} {\bibfield  {journal} {\bibinfo
  {journal} {Rev. Mod. Phys.}\ }\textbf {\bibinfo {volume} {75}},\ \bibinfo
  {pages} {1201} (\bibinfo {year} {2003})}\BibitemShut {NoStop}%
\bibitem [{\citenamefont {Huang}\ \emph {et~al.}(2017)\citenamefont {Huang},
  \citenamefont {Mendl}, \citenamefont {Liu}, \citenamefont {Johnston},
  \citenamefont {Jiang}, \citenamefont {Moritz},\ and\ \citenamefont
  {Devereaux}}]{huang2017}%
  \BibitemOpen
  \bibfield  {author} {\bibinfo {author} {\bibfnamefont {E.~W.}\ \bibnamefont
  {Huang}}, \bibinfo {author} {\bibfnamefont {C.~B.}\ \bibnamefont {Mendl}},
  \bibinfo {author} {\bibfnamefont {S.}~\bibnamefont {Liu}}, \bibinfo {author}
  {\bibfnamefont {S.}~\bibnamefont {Johnston}}, \bibinfo {author}
  {\bibfnamefont {H.-C.}\ \bibnamefont {Jiang}}, \bibinfo {author}
  {\bibfnamefont {B.}~\bibnamefont {Moritz}},\ and\ \bibinfo {author}
  {\bibfnamefont {T.~P.}\ \bibnamefont {Devereaux}},\ }\href
  {https://doi.org/10.1126/science.aak9546} {\bibfield  {journal} {\bibinfo
  {journal} {Science}\ }\textbf {\bibinfo {volume} {358}},\ \bibinfo {pages}
  {1161} (\bibinfo {year} {2017})}\BibitemShut {NoStop}%
\bibitem [{\citenamefont {Zheng}\ \emph {et~al.}(2017)\citenamefont {Zheng},
  \citenamefont {Chung}, \citenamefont {Corboz}, \citenamefont {Ehlers},
  \citenamefont {Qin}, \citenamefont {Noack}, \citenamefont {Shi},
  \citenamefont {White}, \citenamefont {Zhang},\ and\ \citenamefont
  {Chan}}]{zheng2017}%
  \BibitemOpen
  \bibfield  {author} {\bibinfo {author} {\bibfnamefont {B.-X.}\ \bibnamefont
  {Zheng}}, \bibinfo {author} {\bibfnamefont {C.-M.}\ \bibnamefont {Chung}},
  \bibinfo {author} {\bibfnamefont {P.}~\bibnamefont {Corboz}}, \bibinfo
  {author} {\bibfnamefont {G.}~\bibnamefont {Ehlers}}, \bibinfo {author}
  {\bibfnamefont {M.-P.}\ \bibnamefont {Qin}}, \bibinfo {author} {\bibfnamefont
  {R.~M.}\ \bibnamefont {Noack}}, \bibinfo {author} {\bibfnamefont
  {H.}~\bibnamefont {Shi}}, \bibinfo {author} {\bibfnamefont {S.~R.}\
  \bibnamefont {White}}, \bibinfo {author} {\bibfnamefont {S.}~\bibnamefont
  {Zhang}},\ and\ \bibinfo {author} {\bibfnamefont {G.~K.-L.}\ \bibnamefont
  {Chan}},\ }\href {https://doi.org/10.1126/science.aam7127} {\bibfield
  {journal} {\bibinfo  {journal} {Science}\ }\textbf {\bibinfo {volume}
  {358}},\ \bibinfo {pages} {1155} (\bibinfo {year} {2017})}\BibitemShut
  {NoStop}%
\bibitem [{\citenamefont {Huang}\ \emph {et~al.}(2018)\citenamefont {Huang},
  \citenamefont {Mendl}, \citenamefont {Jiang}, \citenamefont {Moritz},\ and\
  \citenamefont {Devereaux}}]{Huang2018}%
  \BibitemOpen
  \bibfield  {author} {\bibinfo {author} {\bibfnamefont {E.~W.}\ \bibnamefont
  {Huang}}, \bibinfo {author} {\bibfnamefont {C.~B.}\ \bibnamefont {Mendl}},
  \bibinfo {author} {\bibfnamefont {H.-C.}\ \bibnamefont {Jiang}}, \bibinfo
  {author} {\bibfnamefont {B.}~\bibnamefont {Moritz}},\ and\ \bibinfo {author}
  {\bibfnamefont {T.~P.}\ \bibnamefont {Devereaux}},\ }\href
  {https://doi.org/10.1038/s41535-018-0097-0} {\bibfield  {journal} {\bibinfo
  {journal} {npj Quantum Materials}\ }\textbf {\bibinfo {volume} {3}},\
  \bibinfo {pages} {22} (\bibinfo {year} {2018})}\BibitemShut {NoStop}%
\bibitem [{\citenamefont {Tranquada}\ \emph {et~al.}(1995)\citenamefont
  {Tranquada}, \citenamefont {Sternlieb}, \citenamefont {Axe}, \citenamefont
  {Nakamura},\ and\ \citenamefont {Uchida}}]{tranquada1995}%
  \BibitemOpen
  \bibfield  {author} {\bibinfo {author} {\bibfnamefont {J.~M.}\ \bibnamefont
  {Tranquada}}, \bibinfo {author} {\bibfnamefont {B.~J.}\ \bibnamefont
  {Sternlieb}}, \bibinfo {author} {\bibfnamefont {J.~D.}\ \bibnamefont {Axe}},
  \bibinfo {author} {\bibfnamefont {Y.}~\bibnamefont {Nakamura}},\ and\
  \bibinfo {author} {\bibfnamefont {S.}~\bibnamefont {Uchida}},\ }\href
  {https://doi.org/10.1038/375561a0} {\bibfield  {journal} {\bibinfo  {journal}
  {Nature}\ }\textbf {\bibinfo {volume} {375}},\ \bibinfo {pages} {561}
  (\bibinfo {year} {1995})}\BibitemShut {NoStop}%
\bibitem [{\citenamefont {Tranquada}\ \emph {et~al.}(2021)\citenamefont
  {Tranquada}, \citenamefont {Dean},\ and\ \citenamefont {Li}}]{tranquada2021}%
  \BibitemOpen
  \bibfield  {author} {\bibinfo {author} {\bibfnamefont {J.~M.}\ \bibnamefont
  {Tranquada}}, \bibinfo {author} {\bibfnamefont {M.~P.~M.}\ \bibnamefont
  {Dean}},\ and\ \bibinfo {author} {\bibfnamefont {Q.}~\bibnamefont {Li}},\
  }\href {https://doi.org/10.7566/JPSJ.90.111002} {\bibfield  {journal}
  {\bibinfo  {journal} {Journal of the Physical Society of Japan}\ }\textbf
  {\bibinfo {volume} {90}},\ \bibinfo {pages} {111002} (\bibinfo {year}
  {2021})}\BibitemShut {NoStop}%
\bibitem [{\citenamefont {Feng}\ \emph {et~al.}(2006)\citenamefont {Feng},
  \citenamefont {Shen}, \citenamefont {Zhou}, \citenamefont {Shen},
  \citenamefont {Lu},\ and\ \citenamefont {Marel}}]{feng2006}%
  \BibitemOpen
  \bibfield  {author} {\bibinfo {author} {\bibfnamefont {D.}~\bibnamefont
  {Feng}}, \bibinfo {author} {\bibfnamefont {Z.-X.}\ \bibnamefont {Shen}},
  \bibinfo {author} {\bibfnamefont {X.}~\bibnamefont {Zhou}}, \bibinfo {author}
  {\bibfnamefont {K.}~\bibnamefont {Shen}}, \bibinfo {author} {\bibfnamefont
  {D.}~\bibnamefont {Lu}},\ and\ \bibinfo {author} {\bibfnamefont
  {D.}~\bibnamefont {Marel}},\ }\href
  {https://doi.org/https://doi.org/10.1016/j.jpcs.2005.10.149} {\bibfield
  {journal} {\bibinfo  {journal} {Journal of Physics and Chemistry of Solids}\
  }\textbf {\bibinfo {volume} {67}},\ \bibinfo {pages} {198} (\bibinfo {year}
  {2006})},\ \bibinfo {note} {spectroscopies in Novel Superconductors
  2004}\BibitemShut {NoStop}%
\bibitem [{\citenamefont {Corboz}\ \emph {et~al.}(2014)\citenamefont {Corboz},
  \citenamefont {Rice},\ and\ \citenamefont {Troyer}}]{Corboz2014}%
  \BibitemOpen
  \bibfield  {author} {\bibinfo {author} {\bibfnamefont {P.}~\bibnamefont
  {Corboz}}, \bibinfo {author} {\bibfnamefont {T.~M.}\ \bibnamefont {Rice}},\
  and\ \bibinfo {author} {\bibfnamefont {M.}~\bibnamefont {Troyer}},\ }\href
  {https://doi.org/10.1103/PhysRevLett.113.046402} {\bibfield  {journal}
  {\bibinfo  {journal} {Phys. Rev. Lett.}\ }\textbf {\bibinfo {volume} {113}},\
  \bibinfo {pages} {046402} (\bibinfo {year} {2014})}\BibitemShut {NoStop}%
\bibitem [{\citenamefont {Arovas}\ \emph {et~al.}(2022)\citenamefont {Arovas},
  \citenamefont {Berg}, \citenamefont {Kivelson},\ and\ \citenamefont
  {Raghu}}]{arovas2022}%
  \BibitemOpen
  \bibfield  {author} {\bibinfo {author} {\bibfnamefont {D.~P.}\ \bibnamefont
  {Arovas}}, \bibinfo {author} {\bibfnamefont {E.}~\bibnamefont {Berg}},
  \bibinfo {author} {\bibfnamefont {S.~A.}\ \bibnamefont {Kivelson}},\ and\
  \bibinfo {author} {\bibfnamefont {S.}~\bibnamefont {Raghu}},\ }\href
  {https://doi.org/10.1146/annurev-conmatphys-031620-102024} {\bibfield
  {journal} {\bibinfo  {journal} {Annual Review of Condensed Matter Physics}\
  }\textbf {\bibinfo {volume} {13}},\ \bibinfo {pages} {null} (\bibinfo {year}
  {2022})}\BibitemShut {NoStop}%
\bibitem [{\citenamefont {Qin}\ \emph {et~al.}(2022)\citenamefont {Qin},
  \citenamefont {Schäfer}, \citenamefont {Andergassen}, \citenamefont
  {Corboz},\ and\ \citenamefont {Gull}}]{qin2022}%
  \BibitemOpen
  \bibfield  {author} {\bibinfo {author} {\bibfnamefont {M.}~\bibnamefont
  {Qin}}, \bibinfo {author} {\bibfnamefont {T.}~\bibnamefont {Schäfer}},
  \bibinfo {author} {\bibfnamefont {S.}~\bibnamefont {Andergassen}}, \bibinfo
  {author} {\bibfnamefont {P.}~\bibnamefont {Corboz}},\ and\ \bibinfo {author}
  {\bibfnamefont {E.}~\bibnamefont {Gull}},\ }\href
  {https://doi.org/10.1146/annurev-conmatphys-090921-033948} {\bibfield
  {journal} {\bibinfo  {journal} {Annual Review of Condensed Matter Physics}\
  }\textbf {\bibinfo {volume} {13}},\ \bibinfo {pages} {null} (\bibinfo {year}
  {2022})}\BibitemShut {NoStop}%
\bibitem [{\citenamefont {Jiang}\ and\ \citenamefont
  {Devereaux}(2019)}]{jiang2019}%
  \BibitemOpen
  \bibfield  {author} {\bibinfo {author} {\bibfnamefont {H.-C.}\ \bibnamefont
  {Jiang}}\ and\ \bibinfo {author} {\bibfnamefont {T.~P.}\ \bibnamefont
  {Devereaux}},\ }\href {https://doi.org/10.1126/science.aal5304} {\bibfield
  {journal} {\bibinfo  {journal} {Science}\ }\textbf {\bibinfo {volume}
  {365}},\ \bibinfo {pages} {1424} (\bibinfo {year} {2019})}\BibitemShut
  {NoStop}%
\bibitem [{\citenamefont {Qin}\ \emph {et~al.}(2020)\citenamefont {Qin},
  \citenamefont {Chung}, \citenamefont {Shi}, \citenamefont {Vitali},
  \citenamefont {Hubig}, \citenamefont {Schollw\"ock}, \citenamefont {White},\
  and\ \citenamefont {Zhang}}]{qin2020}%
  \BibitemOpen
  \bibfield  {author} {\bibinfo {author} {\bibfnamefont {M.}~\bibnamefont
  {Qin}}, \bibinfo {author} {\bibfnamefont {C.-M.}\ \bibnamefont {Chung}},
  \bibinfo {author} {\bibfnamefont {H.}~\bibnamefont {Shi}}, \bibinfo {author}
  {\bibfnamefont {E.}~\bibnamefont {Vitali}}, \bibinfo {author} {\bibfnamefont
  {C.}~\bibnamefont {Hubig}}, \bibinfo {author} {\bibfnamefont
  {U.}~\bibnamefont {Schollw\"ock}}, \bibinfo {author} {\bibfnamefont {S.~R.}\
  \bibnamefont {White}},\ and\ \bibinfo {author} {\bibfnamefont
  {S.}~\bibnamefont {Zhang}} (\bibinfo {collaboration} {Simons Collaboration on
  the Many-Electron Problem}),\ }\href
  {https://doi.org/10.1103/PhysRevX.10.031016} {\bibfield  {journal} {\bibinfo
  {journal} {Phys. Rev. X}\ }\textbf {\bibinfo {volume} {10}},\ \bibinfo
  {pages} {031016} (\bibinfo {year} {2020})}\BibitemShut {NoStop}%
\bibitem [{\citenamefont {Chung}\ \emph {et~al.}(2020)\citenamefont {Chung},
  \citenamefont {Qin}, \citenamefont {Zhang}, \citenamefont {Schollw\"ock},\
  and\ \citenamefont {White}}]{Chung2020}%
  \BibitemOpen
  \bibfield  {author} {\bibinfo {author} {\bibfnamefont {C.-M.}\ \bibnamefont
  {Chung}}, \bibinfo {author} {\bibfnamefont {M.}~\bibnamefont {Qin}}, \bibinfo
  {author} {\bibfnamefont {S.}~\bibnamefont {Zhang}}, \bibinfo {author}
  {\bibfnamefont {U.}~\bibnamefont {Schollw\"ock}},\ and\ \bibinfo {author}
  {\bibfnamefont {S.~R.}\ \bibnamefont {White}} (\bibinfo {collaboration} {The
  Simons Collaboration on the Many-Electron Problem}),\ }\href
  {https://doi.org/10.1103/PhysRevB.102.041106} {\bibfield  {journal} {\bibinfo
   {journal} {Phys. Rev. B}\ }\textbf {\bibinfo {volume} {102}},\ \bibinfo
  {pages} {041106} (\bibinfo {year} {2020})}\BibitemShut {NoStop}%
\bibitem [{\citenamefont {Jiang}\ \emph {et~al.}(2020)\citenamefont {Jiang},
  \citenamefont {Zaanen}, \citenamefont {Devereaux},\ and\ \citenamefont
  {Jiang}}]{jiang2020}%
  \BibitemOpen
  \bibfield  {author} {\bibinfo {author} {\bibfnamefont {Y.-F.}\ \bibnamefont
  {Jiang}}, \bibinfo {author} {\bibfnamefont {J.}~\bibnamefont {Zaanen}},
  \bibinfo {author} {\bibfnamefont {T.~P.}\ \bibnamefont {Devereaux}},\ and\
  \bibinfo {author} {\bibfnamefont {H.-C.}\ \bibnamefont {Jiang}},\ }\href
  {https://doi.org/10.1103/PhysRevResearch.2.033073} {\bibfield  {journal}
  {\bibinfo  {journal} {Phys. Rev. Research}\ }\textbf {\bibinfo {volume}
  {2}},\ \bibinfo {pages} {033073} (\bibinfo {year} {2020})}\BibitemShut
  {NoStop}%
\bibitem [{\citenamefont {Sorella}(2021)}]{sorella2021}%
  \BibitemOpen
  \bibfield  {author} {\bibinfo {author} {\bibfnamefont {S.}~\bibnamefont
  {Sorella}},\ }\href@noop {} {} (\bibinfo {year} {2021}),\ \Eprint
  {https://arxiv.org/abs/2101.07045} {arXiv:2101.07045 [cond-mat.str-el]}
  \BibitemShut {NoStop}%
\bibitem [{\citenamefont {Wietek}\ \emph {et~al.}(2021)\citenamefont {Wietek},
  \citenamefont {He}, \citenamefont {White}, \citenamefont {Georges},\ and\
  \citenamefont {Stoudenmire}}]{wietek2021}%
  \BibitemOpen
  \bibfield  {author} {\bibinfo {author} {\bibfnamefont {A.}~\bibnamefont
  {Wietek}}, \bibinfo {author} {\bibfnamefont {Y.-Y.}\ \bibnamefont {He}},
  \bibinfo {author} {\bibfnamefont {S.~R.}\ \bibnamefont {White}}, \bibinfo
  {author} {\bibfnamefont {A.}~\bibnamefont {Georges}},\ and\ \bibinfo {author}
  {\bibfnamefont {E.~M.}\ \bibnamefont {Stoudenmire}},\ }\href
  {https://doi.org/10.1103/physrevx.11.031007} {\bibfield  {journal} {\bibinfo
  {journal} {Physical Review X}\ }\textbf {\bibinfo {volume} {11}},\ \bibinfo
  {pages} {031007} (\bibinfo {year} {2021})}\BibitemShut {NoStop}%
\bibitem [{\citenamefont {Mai}\ \emph {et~al.}(2022)\citenamefont {Mai},
  \citenamefont {Karakuzu}, \citenamefont {Balduzzi}, \citenamefont
  {Johnston},\ and\ \citenamefont {Maier}}]{mai2021}%
  \BibitemOpen
  \bibfield  {author} {\bibinfo {author} {\bibfnamefont {P.}~\bibnamefont
  {Mai}}, \bibinfo {author} {\bibfnamefont {S.}~\bibnamefont {Karakuzu}},
  \bibinfo {author} {\bibfnamefont {G.}~\bibnamefont {Balduzzi}}, \bibinfo
  {author} {\bibfnamefont {S.}~\bibnamefont {Johnston}},\ and\ \bibinfo
  {author} {\bibfnamefont {T.~A.}\ \bibnamefont {Maier}},\ }\href
  {https://doi.org/10.1073/pnas.2112806119} {\bibfield  {journal} {\bibinfo
  {journal} {Proceedings of the National Academy of Sciences}\ }\textbf
  {\bibinfo {volume} {119}},\ \bibinfo {pages} {e2112806119} (\bibinfo {year}
  {2022})}\BibitemShut {NoStop}%
\bibitem [{\citenamefont {Brown}\ \emph {et~al.}(2019)\citenamefont {Brown},
  \citenamefont {Mitra}, \citenamefont {Guardado-Sanchez}, \citenamefont
  {Nourafkan}, \citenamefont {Reymbaut}, \citenamefont {Hébert}, \citenamefont
  {Bergeron}, \citenamefont {Tremblay}, \citenamefont {Kokalj}, \citenamefont
  {Huse}, \citenamefont {Schauß},\ and\ \citenamefont {Bakr}}]{brown2019}%
  \BibitemOpen
  \bibfield  {author} {\bibinfo {author} {\bibfnamefont {P.~T.}\ \bibnamefont
  {Brown}}, \bibinfo {author} {\bibfnamefont {D.}~\bibnamefont {Mitra}},
  \bibinfo {author} {\bibfnamefont {E.}~\bibnamefont {Guardado-Sanchez}},
  \bibinfo {author} {\bibfnamefont {R.}~\bibnamefont {Nourafkan}}, \bibinfo
  {author} {\bibfnamefont {A.}~\bibnamefont {Reymbaut}}, \bibinfo {author}
  {\bibfnamefont {C.-D.}\ \bibnamefont {Hébert}}, \bibinfo {author}
  {\bibfnamefont {S.}~\bibnamefont {Bergeron}}, \bibinfo {author}
  {\bibfnamefont {A.-M.~S.}\ \bibnamefont {Tremblay}}, \bibinfo {author}
  {\bibfnamefont {J.}~\bibnamefont {Kokalj}}, \bibinfo {author} {\bibfnamefont
  {D.~A.}\ \bibnamefont {Huse}}, \bibinfo {author} {\bibfnamefont
  {P.}~\bibnamefont {Schauß}},\ and\ \bibinfo {author} {\bibfnamefont {W.~S.}\
  \bibnamefont {Bakr}},\ }\href {https://doi.org/10.1126/science.aat4134}
  {\bibfield  {journal} {\bibinfo  {journal} {Science}\ }\textbf {\bibinfo
  {volume} {363}},\ \bibinfo {pages} {379} (\bibinfo {year}
  {2019})}\BibitemShut {NoStop}%
\bibitem [{\citenamefont {Huang}\ \emph {et~al.}(2019)\citenamefont {Huang},
  \citenamefont {Sheppard}, \citenamefont {Moritz},\ and\ \citenamefont
  {Devereaux}}]{huang2019}%
  \BibitemOpen
  \bibfield  {author} {\bibinfo {author} {\bibfnamefont {E.~W.}\ \bibnamefont
  {Huang}}, \bibinfo {author} {\bibfnamefont {R.}~\bibnamefont {Sheppard}},
  \bibinfo {author} {\bibfnamefont {B.}~\bibnamefont {Moritz}},\ and\ \bibinfo
  {author} {\bibfnamefont {T.~P.}\ \bibnamefont {Devereaux}},\ }\href
  {https://doi.org/10.1126/science.aau7063} {\bibfield  {journal} {\bibinfo
  {journal} {Science}\ }\textbf {\bibinfo {volume} {366}},\ \bibinfo {pages}
  {987–990} (\bibinfo {year} {2019})}\BibitemShut {NoStop}%
\bibitem [{\citenamefont {Arpaia}\ \emph {et~al.}(2019)\citenamefont {Arpaia},
  \citenamefont {Caprara}, \citenamefont {Fumagalli}, \citenamefont {Vecchi},
  \citenamefont {Peng}, \citenamefont {Andersson}, \citenamefont {Betto},
  \citenamefont {Luca}, \citenamefont {Brookes}, \citenamefont {Lombardi},
  \citenamefont {Salluzzo}, \citenamefont {Braicovich}, \citenamefont {Castro},
  \citenamefont {Grilli},\ and\ \citenamefont {Ghiringhelli}}]{arpaia2019}%
  \BibitemOpen
  \bibfield  {author} {\bibinfo {author} {\bibfnamefont {R.}~\bibnamefont
  {Arpaia}}, \bibinfo {author} {\bibfnamefont {S.}~\bibnamefont {Caprara}},
  \bibinfo {author} {\bibfnamefont {R.}~\bibnamefont {Fumagalli}}, \bibinfo
  {author} {\bibfnamefont {G.~D.}\ \bibnamefont {Vecchi}}, \bibinfo {author}
  {\bibfnamefont {Y.~Y.}\ \bibnamefont {Peng}}, \bibinfo {author}
  {\bibfnamefont {E.}~\bibnamefont {Andersson}}, \bibinfo {author}
  {\bibfnamefont {D.}~\bibnamefont {Betto}}, \bibinfo {author} {\bibfnamefont
  {G.~M.~D.}\ \bibnamefont {Luca}}, \bibinfo {author} {\bibfnamefont {N.~B.}\
  \bibnamefont {Brookes}}, \bibinfo {author} {\bibfnamefont {F.}~\bibnamefont
  {Lombardi}}, \bibinfo {author} {\bibfnamefont {M.}~\bibnamefont {Salluzzo}},
  \bibinfo {author} {\bibfnamefont {L.}~\bibnamefont {Braicovich}}, \bibinfo
  {author} {\bibfnamefont {C.~D.}\ \bibnamefont {Castro}}, \bibinfo {author}
  {\bibfnamefont {M.}~\bibnamefont {Grilli}},\ and\ \bibinfo {author}
  {\bibfnamefont {G.}~\bibnamefont {Ghiringhelli}},\ }\href
  {https://doi.org/10.1126/science.aav1315} {\bibfield  {journal} {\bibinfo
  {journal} {Science}\ }\textbf {\bibinfo {volume} {365}},\ \bibinfo {pages}
  {906} (\bibinfo {year} {2019})}\BibitemShut {NoStop}%
\bibitem [{\citenamefont {Miao}\ \emph {et~al.}(2021)\citenamefont {Miao},
  \citenamefont {Fabbris}, \citenamefont {Koch}, \citenamefont {Mazzone},
  \citenamefont {Nelson}, \citenamefont {Acevedo-Esteves}, \citenamefont {Gu},
  \citenamefont {Li}, \citenamefont {Yilimaz}, \citenamefont {Kaznatcheev},
  \citenamefont {Vescovo}, \citenamefont {Oda}, \citenamefont {Kurosawa},
  \citenamefont {Momono}, \citenamefont {Assefa}, \citenamefont {Robinson},
  \citenamefont {Bozin}, \citenamefont {Tranquada}, \citenamefont {Johnson},\
  and\ \citenamefont {Dean}}]{miao2021}%
  \BibitemOpen
  \bibfield  {author} {\bibinfo {author} {\bibfnamefont {H.}~\bibnamefont
  {Miao}}, \bibinfo {author} {\bibfnamefont {G.}~\bibnamefont {Fabbris}},
  \bibinfo {author} {\bibfnamefont {R.~J.}\ \bibnamefont {Koch}}, \bibinfo
  {author} {\bibfnamefont {D.~G.}\ \bibnamefont {Mazzone}}, \bibinfo {author}
  {\bibfnamefont {C.~S.}\ \bibnamefont {Nelson}}, \bibinfo {author}
  {\bibfnamefont {R.}~\bibnamefont {Acevedo-Esteves}}, \bibinfo {author}
  {\bibfnamefont {G.~D.}\ \bibnamefont {Gu}}, \bibinfo {author} {\bibfnamefont
  {Y.}~\bibnamefont {Li}}, \bibinfo {author} {\bibfnamefont {T.}~\bibnamefont
  {Yilimaz}}, \bibinfo {author} {\bibfnamefont {K.}~\bibnamefont
  {Kaznatcheev}}, \bibinfo {author} {\bibfnamefont {E.}~\bibnamefont
  {Vescovo}}, \bibinfo {author} {\bibfnamefont {M.}~\bibnamefont {Oda}},
  \bibinfo {author} {\bibfnamefont {T.}~\bibnamefont {Kurosawa}}, \bibinfo
  {author} {\bibfnamefont {N.}~\bibnamefont {Momono}}, \bibinfo {author}
  {\bibfnamefont {T.}~\bibnamefont {Assefa}}, \bibinfo {author} {\bibfnamefont
  {I.~K.}\ \bibnamefont {Robinson}}, \bibinfo {author} {\bibfnamefont {E.~S.}\
  \bibnamefont {Bozin}}, \bibinfo {author} {\bibfnamefont {J.~M.}\ \bibnamefont
  {Tranquada}}, \bibinfo {author} {\bibfnamefont {P.~D.}\ \bibnamefont
  {Johnson}},\ and\ \bibinfo {author} {\bibfnamefont {M.~P.~M.}\ \bibnamefont
  {Dean}},\ }\href {https://doi.org/10.1038/s41535-021-00327-4} {\bibfield
  {journal} {\bibinfo  {journal} {npj Quantum Materials}\ }\textbf {\bibinfo
  {volume} {6}},\ \bibinfo {pages} {31} (\bibinfo {year} {2021})}\BibitemShut
  {NoStop}%
\bibitem [{\citenamefont {Lin}\ \emph {et~al.}(2020)\citenamefont {Lin},
  \citenamefont {Miao}, \citenamefont {Mazzone}, \citenamefont {Gu},
  \citenamefont {Nag}, \citenamefont {Walters}, \citenamefont
  {Garc\'{\i}a-Fern\'andez}, \citenamefont {Barbour}, \citenamefont
  {Pelliciari}, \citenamefont {Jarrige}, \citenamefont {Oda}, \citenamefont
  {Kurosawa}, \citenamefont {Momono}, \citenamefont {Zhou}, \citenamefont
  {Bisogni}, \citenamefont {Liu},\ and\ \citenamefont {Dean}}]{lin2020}%
  \BibitemOpen
  \bibfield  {author} {\bibinfo {author} {\bibfnamefont {J.~Q.}\ \bibnamefont
  {Lin}}, \bibinfo {author} {\bibfnamefont {H.}~\bibnamefont {Miao}}, \bibinfo
  {author} {\bibfnamefont {D.~G.}\ \bibnamefont {Mazzone}}, \bibinfo {author}
  {\bibfnamefont {G.~D.}\ \bibnamefont {Gu}}, \bibinfo {author} {\bibfnamefont
  {A.}~\bibnamefont {Nag}}, \bibinfo {author} {\bibfnamefont {A.~C.}\
  \bibnamefont {Walters}}, \bibinfo {author} {\bibfnamefont {M.}~\bibnamefont
  {Garc\'{\i}a-Fern\'andez}}, \bibinfo {author} {\bibfnamefont
  {A.}~\bibnamefont {Barbour}}, \bibinfo {author} {\bibfnamefont
  {J.}~\bibnamefont {Pelliciari}}, \bibinfo {author} {\bibfnamefont
  {I.}~\bibnamefont {Jarrige}}, \bibinfo {author} {\bibfnamefont
  {M.}~\bibnamefont {Oda}}, \bibinfo {author} {\bibfnamefont {K.}~\bibnamefont
  {Kurosawa}}, \bibinfo {author} {\bibfnamefont {N.}~\bibnamefont {Momono}},
  \bibinfo {author} {\bibfnamefont {K.-J.}\ \bibnamefont {Zhou}}, \bibinfo
  {author} {\bibfnamefont {V.}~\bibnamefont {Bisogni}}, \bibinfo {author}
  {\bibfnamefont {X.}~\bibnamefont {Liu}},\ and\ \bibinfo {author}
  {\bibfnamefont {M.~P.~M.}\ \bibnamefont {Dean}},\ }\href
  {https://doi.org/10.1103/PhysRevLett.124.207005} {\bibfield  {journal}
  {\bibinfo  {journal} {Phys. Rev. Lett.}\ }\textbf {\bibinfo {volume} {124}},\
  \bibinfo {pages} {207005} (\bibinfo {year} {2020})}\BibitemShut {NoStop}%
\bibitem [{\citenamefont {Peng}\ \emph {et~al.}(2018)\citenamefont {Peng},
  \citenamefont {Fumagalli}, \citenamefont {Ding}, \citenamefont {Minola},
  \citenamefont {Caprara}, \citenamefont {Betto}, \citenamefont {Bluschke},
  \citenamefont {De~Luca}, \citenamefont {Kummer}, \citenamefont {Lefrançois},
  \citenamefont {Salluzzo}, \citenamefont {Suzuki}, \citenamefont {Le~Tacon},
  \citenamefont {Zhou}, \citenamefont {Brookes}, \citenamefont {Keimer},
  \citenamefont {Braicovich}, \citenamefont {Grilli},\ and\ \citenamefont
  {Ghiringhelli}}]{peng2018}%
  \BibitemOpen
  \bibfield  {author} {\bibinfo {author} {\bibfnamefont {Y.~Y.}\ \bibnamefont
  {Peng}}, \bibinfo {author} {\bibfnamefont {R.}~\bibnamefont {Fumagalli}},
  \bibinfo {author} {\bibfnamefont {Y.}~\bibnamefont {Ding}}, \bibinfo {author}
  {\bibfnamefont {M.}~\bibnamefont {Minola}}, \bibinfo {author} {\bibfnamefont
  {S.}~\bibnamefont {Caprara}}, \bibinfo {author} {\bibfnamefont
  {D.}~\bibnamefont {Betto}}, \bibinfo {author} {\bibfnamefont
  {M.}~\bibnamefont {Bluschke}}, \bibinfo {author} {\bibfnamefont {G.~M.}\
  \bibnamefont {De~Luca}}, \bibinfo {author} {\bibfnamefont {K.}~\bibnamefont
  {Kummer}}, \bibinfo {author} {\bibfnamefont {E.}~\bibnamefont {Lefrançois}},
  \bibinfo {author} {\bibfnamefont {M.}~\bibnamefont {Salluzzo}}, \bibinfo
  {author} {\bibfnamefont {H.}~\bibnamefont {Suzuki}}, \bibinfo {author}
  {\bibfnamefont {M.}~\bibnamefont {Le~Tacon}}, \bibinfo {author}
  {\bibfnamefont {X.~J.}\ \bibnamefont {Zhou}}, \bibinfo {author}
  {\bibfnamefont {N.~B.}\ \bibnamefont {Brookes}}, \bibinfo {author}
  {\bibfnamefont {B.}~\bibnamefont {Keimer}}, \bibinfo {author} {\bibfnamefont
  {L.}~\bibnamefont {Braicovich}}, \bibinfo {author} {\bibfnamefont
  {M.}~\bibnamefont {Grilli}},\ and\ \bibinfo {author} {\bibfnamefont
  {G.}~\bibnamefont {Ghiringhelli}},\ }\href
  {https://doi.org/10.1038/s41563-018-0108-3} {\bibfield  {journal} {\bibinfo
  {journal} {Nature Materials}\ }\textbf {\bibinfo {volume} {17}},\ \bibinfo
  {pages} {697–702} (\bibinfo {year} {2018})}\BibitemShut {NoStop}%
\bibitem [{\citenamefont {Lee}\ \emph {et~al.}(2020)\citenamefont {Lee},
  \citenamefont {Zhou}, \citenamefont {Hepting}, \citenamefont {Li},
  \citenamefont {Nag}, \citenamefont {Walters}, \citenamefont
  {Garcia-Fernandez}, \citenamefont {Robarts}, \citenamefont {Hashimoto},
  \citenamefont {Lu}, \citenamefont {Nosarzewski}, \citenamefont {Song},
  \citenamefont {Eisaki}, \citenamefont {Shen}, \citenamefont {Moritz},
  \citenamefont {Zaanen},\ and\ \citenamefont {Devereaux}}]{lee2020}%
  \BibitemOpen
  \bibfield  {author} {\bibinfo {author} {\bibfnamefont {W.~S.}\ \bibnamefont
  {Lee}}, \bibinfo {author} {\bibfnamefont {K.-J.}\ \bibnamefont {Zhou}},
  \bibinfo {author} {\bibfnamefont {M.}~\bibnamefont {Hepting}}, \bibinfo
  {author} {\bibfnamefont {J.}~\bibnamefont {Li}}, \bibinfo {author}
  {\bibfnamefont {A.}~\bibnamefont {Nag}}, \bibinfo {author} {\bibfnamefont
  {A.~C.}\ \bibnamefont {Walters}}, \bibinfo {author} {\bibfnamefont
  {M.}~\bibnamefont {Garcia-Fernandez}}, \bibinfo {author} {\bibfnamefont
  {H.~C.}\ \bibnamefont {Robarts}}, \bibinfo {author} {\bibfnamefont
  {M.}~\bibnamefont {Hashimoto}}, \bibinfo {author} {\bibfnamefont
  {H.}~\bibnamefont {Lu}}, \bibinfo {author} {\bibfnamefont {B.}~\bibnamefont
  {Nosarzewski}}, \bibinfo {author} {\bibfnamefont {D.}~\bibnamefont {Song}},
  \bibinfo {author} {\bibfnamefont {H.}~\bibnamefont {Eisaki}}, \bibinfo
  {author} {\bibfnamefont {Z.~X.}\ \bibnamefont {Shen}}, \bibinfo {author}
  {\bibfnamefont {B.}~\bibnamefont {Moritz}}, \bibinfo {author} {\bibfnamefont
  {J.}~\bibnamefont {Zaanen}},\ and\ \bibinfo {author} {\bibfnamefont {T.~P.}\
  \bibnamefont {Devereaux}},\ }\href
  {https://doi.org/10.1038/s41567-020-0993-7} {\bibfield  {journal} {\bibinfo
  {journal} {Nature Physics}\ }\textbf {\bibinfo {volume} {17}},\ \bibinfo
  {pages} {53–57} (\bibinfo {year} {2020})}\BibitemShut {NoStop}%
\bibitem [{\citenamefont {Ma}\ \emph {et~al.}(2021)\citenamefont {Ma},
  \citenamefont {Rule}, \citenamefont {Cronkwright}, \citenamefont {Dragomir},
  \citenamefont {Mitchell}, \citenamefont {Smith}, \citenamefont {Chi},
  \citenamefont {Kolesnikov}, \citenamefont {Stone},\ and\ \citenamefont
  {Gaulin}}]{ma2021}%
  \BibitemOpen
  \bibfield  {author} {\bibinfo {author} {\bibfnamefont {Q.}~\bibnamefont
  {Ma}}, \bibinfo {author} {\bibfnamefont {K.~C.}\ \bibnamefont {Rule}},
  \bibinfo {author} {\bibfnamefont {Z.~W.}\ \bibnamefont {Cronkwright}},
  \bibinfo {author} {\bibfnamefont {M.}~\bibnamefont {Dragomir}}, \bibinfo
  {author} {\bibfnamefont {G.}~\bibnamefont {Mitchell}}, \bibinfo {author}
  {\bibfnamefont {E.~M.}\ \bibnamefont {Smith}}, \bibinfo {author}
  {\bibfnamefont {S.}~\bibnamefont {Chi}}, \bibinfo {author} {\bibfnamefont
  {A.~I.}\ \bibnamefont {Kolesnikov}}, \bibinfo {author} {\bibfnamefont
  {M.~B.}\ \bibnamefont {Stone}},\ and\ \bibinfo {author} {\bibfnamefont
  {B.~D.}\ \bibnamefont {Gaulin}},\ }\href
  {https://doi.org/10.1103/PhysRevResearch.3.023151} {\bibfield  {journal}
  {\bibinfo  {journal} {Phys. Rev. Research}\ }\textbf {\bibinfo {volume}
  {3}},\ \bibinfo {pages} {023151} (\bibinfo {year} {2021})}\BibitemShut
  {NoStop}%
\bibitem [{\citenamefont {Tam}\ \emph {et~al.}(2022)\citenamefont {Tam},
  \citenamefont {Zhu}, \citenamefont {Ayres}, \citenamefont {Kummer},
  \citenamefont {Yakhou-Harris}, \citenamefont {Cooper}, \citenamefont
  {Carrington},\ and\ \citenamefont {Hayden}}]{tam2022}%
  \BibitemOpen
  \bibfield  {author} {\bibinfo {author} {\bibfnamefont {C.~C.}\ \bibnamefont
  {Tam}}, \bibinfo {author} {\bibfnamefont {M.}~\bibnamefont {Zhu}}, \bibinfo
  {author} {\bibfnamefont {J.}~\bibnamefont {Ayres}}, \bibinfo {author}
  {\bibfnamefont {K.}~\bibnamefont {Kummer}}, \bibinfo {author} {\bibfnamefont
  {F.}~\bibnamefont {Yakhou-Harris}}, \bibinfo {author} {\bibfnamefont {J.~R.}\
  \bibnamefont {Cooper}}, \bibinfo {author} {\bibfnamefont {A.}~\bibnamefont
  {Carrington}},\ and\ \bibinfo {author} {\bibfnamefont {S.~M.}\ \bibnamefont
  {Hayden}},\ }\href {https://doi.org/10.1038/s41467-022-28124-y} {\bibfield
  {journal} {\bibinfo  {journal} {Nature Communications}\ }\textbf {\bibinfo
  {volume} {13}},\ \bibinfo {pages} {570} (\bibinfo {year} {2022})}\BibitemShut
  {NoStop}%
\bibitem [{\citenamefont {Lee}\ \emph {et~al.}(2021)\citenamefont {Lee},
  \citenamefont {Huang}, \citenamefont {Johnson}, \citenamefont {Guo},
  \citenamefont {Husain}, \citenamefont {Mitrano}, \citenamefont {Lu},
  \citenamefont {Zakrzewski}, \citenamefont {de~la Pe\~{n}a}, \citenamefont
  {Peng}, \citenamefont {Lee}, \citenamefont {Jang}, \citenamefont {Lee},
  \citenamefont {Joe}, \citenamefont {Dorisese}, \citenamefont {Szypryt},
  \citenamefont {Swetz}, \citenamefont {Aczel}, \citenamefont {Macdougall},
  \citenamefont {Kivelson}, \citenamefont {Fradkin},\ and\ \citenamefont
  {Abbamonte}}]{lee2021}%
  \BibitemOpen
  \bibfield  {author} {\bibinfo {author} {\bibfnamefont {S.}~\bibnamefont
  {Lee}}, \bibinfo {author} {\bibfnamefont {E.~W.}\ \bibnamefont {Huang}},
  \bibinfo {author} {\bibfnamefont {T.~A.}\ \bibnamefont {Johnson}}, \bibinfo
  {author} {\bibfnamefont {X.}~\bibnamefont {Guo}}, \bibinfo {author}
  {\bibfnamefont {A.~A.}\ \bibnamefont {Husain}}, \bibinfo {author}
  {\bibfnamefont {M.}~\bibnamefont {Mitrano}}, \bibinfo {author} {\bibfnamefont
  {K.}~\bibnamefont {Lu}}, \bibinfo {author} {\bibfnamefont {A.~V.}\
  \bibnamefont {Zakrzewski}}, \bibinfo {author} {\bibfnamefont
  {G.}~\bibnamefont {de~la Pe\~{n}a}}, \bibinfo {author} {\bibfnamefont
  {Y.}~\bibnamefont {Peng}}, \bibinfo {author} {\bibfnamefont {S.-J.}\
  \bibnamefont {Lee}}, \bibinfo {author} {\bibfnamefont {H.}~\bibnamefont
  {Jang}}, \bibinfo {author} {\bibfnamefont {J.-S.}\ \bibnamefont {Lee}},
  \bibinfo {author} {\bibfnamefont {Y.~I.}\ \bibnamefont {Joe}}, \bibinfo
  {author} {\bibfnamefont {W.~B.}\ \bibnamefont {Dorisese}}, \bibinfo {author}
  {\bibfnamefont {P.}~\bibnamefont {Szypryt}}, \bibinfo {author} {\bibfnamefont
  {D.~S.}\ \bibnamefont {Swetz}}, \bibinfo {author} {\bibfnamefont {A.~A.}\
  \bibnamefont {Aczel}}, \bibinfo {author} {\bibfnamefont {G.~J.}\ \bibnamefont
  {Macdougall}}, \bibinfo {author} {\bibfnamefont {S.~A.}\ \bibnamefont
  {Kivelson}}, \bibinfo {author} {\bibfnamefont {E.}~\bibnamefont {Fradkin}},\
  and\ \bibinfo {author} {\bibfnamefont {P.}~\bibnamefont {Abbamonte}},\ }\href
  {https://arxiv.org/abs/2110.13991} {\bibfield  {journal} {\bibinfo  {journal}
  {arXiv:2110.13991}\ } (\bibinfo {year} {2021})}\BibitemShut {NoStop}%
\bibitem [{\citenamefont {Kawasaki}\ \emph {et~al.}(2021)\citenamefont
  {Kawasaki}, \citenamefont {Ito}, \citenamefont {Kamijima}, \citenamefont
  {Lin},\ and\ \citenamefont {Zheng}}]{kawasaki2021}%
  \BibitemOpen
  \bibfield  {author} {\bibinfo {author} {\bibfnamefont {S.}~\bibnamefont
  {Kawasaki}}, \bibinfo {author} {\bibfnamefont {M.}~\bibnamefont {Ito}},
  \bibinfo {author} {\bibfnamefont {D.}~\bibnamefont {Kamijima}}, \bibinfo
  {author} {\bibfnamefont {C.}~\bibnamefont {Lin}},\ and\ \bibinfo {author}
  {\bibfnamefont {G.-q.}\ \bibnamefont {Zheng}},\ }\href
  {https://doi.org/10.7566/JPSJ.90.111008} {\bibfield  {journal} {\bibinfo
  {journal} {Journal of the Physical Society of Japan}\ }\textbf {\bibinfo
  {volume} {90}},\ \bibinfo {pages} {111008} (\bibinfo {year}
  {2021})}\BibitemShut {NoStop}%
\bibitem [{\citenamefont {White}\ \emph {et~al.}(1989)\citenamefont {White},
  \citenamefont {Scalapino}, \citenamefont {Sugar}, \citenamefont {Loh},
  \citenamefont {Gubernatis},\ and\ \citenamefont {Scalettar}}]{white1989}%
  \BibitemOpen
  \bibfield  {author} {\bibinfo {author} {\bibfnamefont {S.~R.}\ \bibnamefont
  {White}}, \bibinfo {author} {\bibfnamefont {D.~J.}\ \bibnamefont
  {Scalapino}}, \bibinfo {author} {\bibfnamefont {R.~L.}\ \bibnamefont
  {Sugar}}, \bibinfo {author} {\bibfnamefont {E.~Y.}\ \bibnamefont {Loh}},
  \bibinfo {author} {\bibfnamefont {J.~E.}\ \bibnamefont {Gubernatis}},\ and\
  \bibinfo {author} {\bibfnamefont {R.~T.}\ \bibnamefont {Scalettar}},\ }\href
  {https://doi.org/10.1103/PhysRevB.40.506} {\bibfield  {journal} {\bibinfo
  {journal} {Phys. Rev. B}\ }\textbf {\bibinfo {volume} {40}},\ \bibinfo
  {pages} {506} (\bibinfo {year} {1989})}\BibitemShut {NoStop}%
\bibitem [{\citenamefont {Blankenbecler}\ \emph {et~al.}(1981)\citenamefont
  {Blankenbecler}, \citenamefont {Scalapino},\ and\ \citenamefont
  {Sugar}}]{bss1981}%
  \BibitemOpen
  \bibfield  {author} {\bibinfo {author} {\bibfnamefont {R.}~\bibnamefont
  {Blankenbecler}}, \bibinfo {author} {\bibfnamefont {D.~J.}\ \bibnamefont
  {Scalapino}},\ and\ \bibinfo {author} {\bibfnamefont {R.~L.}\ \bibnamefont
  {Sugar}},\ }\href {https://doi.org/10.1103/PhysRevD.24.2278} {\bibfield
  {journal} {\bibinfo  {journal} {Phys. Rev. D}\ }\textbf {\bibinfo {volume}
  {24}},\ \bibinfo {pages} {2278} (\bibinfo {year} {1981})}\BibitemShut
  {NoStop}%
\bibitem [{\citenamefont {Zaanen}\ and\ \citenamefont
  {Gunnarsson}(1989)}]{zaanen1989}%
  \BibitemOpen
  \bibfield  {author} {\bibinfo {author} {\bibfnamefont {J.}~\bibnamefont
  {Zaanen}}\ and\ \bibinfo {author} {\bibfnamefont {O.}~\bibnamefont
  {Gunnarsson}},\ }\href {https://doi.org/10.1103/PhysRevB.40.7391} {\bibfield
  {journal} {\bibinfo  {journal} {Phys. Rev. B}\ }\textbf {\bibinfo {volume}
  {40}},\ \bibinfo {pages} {7391} (\bibinfo {year} {1989})}\BibitemShut
  {NoStop}%
\bibitem [{\citenamefont {Zaanen}\ \emph {et~al.}(2001)\citenamefont {Zaanen},
  \citenamefont {Osman}, \citenamefont {Kruis}, \citenamefont {Nussinov},\ and\
  \citenamefont {Tworzydlo}}]{zaanen2001}%
  \BibitemOpen
  \bibfield  {author} {\bibinfo {author} {\bibfnamefont {J.}~\bibnamefont
  {Zaanen}}, \bibinfo {author} {\bibfnamefont {O.~Y.}\ \bibnamefont {Osman}},
  \bibinfo {author} {\bibfnamefont {H.~V.}\ \bibnamefont {Kruis}}, \bibinfo
  {author} {\bibfnamefont {Z.}~\bibnamefont {Nussinov}},\ and\ \bibinfo
  {author} {\bibfnamefont {J.}~\bibnamefont {Tworzydlo}},\ }\href
  {https://doi.org/10.1080/13642810110051791} {\bibfield  {journal} {\bibinfo
  {journal} {Philosophical Magazine B}\ }\textbf {\bibinfo {volume} {81}},\
  \bibinfo {pages} {1485–1531} (\bibinfo {year} {2001})}\BibitemShut
  {NoStop}%
\bibitem [{\citenamefont {Mondaini}\ \emph {et~al.}(2012)\citenamefont
  {Mondaini}, \citenamefont {Ying}, \citenamefont {Paiva},\ and\ \citenamefont
  {Scalettar}}]{mondaini2012}%
  \BibitemOpen
  \bibfield  {author} {\bibinfo {author} {\bibfnamefont {R.}~\bibnamefont
  {Mondaini}}, \bibinfo {author} {\bibfnamefont {T.}~\bibnamefont {Ying}},
  \bibinfo {author} {\bibfnamefont {T.}~\bibnamefont {Paiva}},\ and\ \bibinfo
  {author} {\bibfnamefont {R.~T.}\ \bibnamefont {Scalettar}},\ }\href
  {https://doi.org/10.1103/PhysRevB.86.184506} {\bibfield  {journal} {\bibinfo
  {journal} {Phys. Rev. B}\ }\textbf {\bibinfo {volume} {86}},\ \bibinfo
  {pages} {184506} (\bibinfo {year} {2012})}\BibitemShut {NoStop}%
\bibitem [{\citenamefont {Liu}\ \emph {et~al.}(2021)\citenamefont {Liu},
  \citenamefont {Jost}, \citenamefont {Moritz}, \citenamefont {Huang},
  \citenamefont {Hackl},\ and\ \citenamefont {Devereaux}}]{liu2021}%
  \BibitemOpen
  \bibfield  {author} {\bibinfo {author} {\bibfnamefont {T.}~\bibnamefont
  {Liu}}, \bibinfo {author} {\bibfnamefont {D.}~\bibnamefont {Jost}}, \bibinfo
  {author} {\bibfnamefont {B.}~\bibnamefont {Moritz}}, \bibinfo {author}
  {\bibfnamefont {E.~W.}\ \bibnamefont {Huang}}, \bibinfo {author}
  {\bibfnamefont {R.}~\bibnamefont {Hackl}},\ and\ \bibinfo {author}
  {\bibfnamefont {T.~P.}\ \bibnamefont {Devereaux}},\ }\href
  {https://doi.org/10.1103/PhysRevB.103.134502} {\bibfield  {journal} {\bibinfo
   {journal} {Phys. Rev. B}\ }\textbf {\bibinfo {volume} {103}},\ \bibinfo
  {pages} {134502} (\bibinfo {year} {2021})}\BibitemShut {NoStop}%
\bibitem [{\citenamefont {Delacrétaz}\ \emph {et~al.}(2017)\citenamefont
  {Delacrétaz}, \citenamefont {Goutéraux}, \citenamefont {Hartnoll},\ and\
  \citenamefont {Karlsson}}]{luca2017scipost}%
  \BibitemOpen
  \bibfield  {author} {\bibinfo {author} {\bibfnamefont {L.~V.}\ \bibnamefont
  {Delacrétaz}}, \bibinfo {author} {\bibfnamefont {B.}~\bibnamefont
  {Goutéraux}}, \bibinfo {author} {\bibfnamefont {S.~A.}\ \bibnamefont
  {Hartnoll}},\ and\ \bibinfo {author} {\bibfnamefont {A.}~\bibnamefont
  {Karlsson}},\ }\href {https://doi.org/10.21468/SciPostPhys.3.3.025}
  {\bibfield  {journal} {\bibinfo  {journal} {SciPost Phys.}\ }\textbf
  {\bibinfo {volume} {3}},\ \bibinfo {pages} {025} (\bibinfo {year}
  {2017})}\BibitemShut {NoStop}%
\bibitem [{\citenamefont {Delacr\'etaz}\ \emph {et~al.}(2017)\citenamefont
  {Delacr\'etaz}, \citenamefont {Gout\'eraux}, \citenamefont {Hartnoll},\ and\
  \citenamefont {Karlsson}}]{luca2017prb}%
  \BibitemOpen
  \bibfield  {author} {\bibinfo {author} {\bibfnamefont {L.~V.}\ \bibnamefont
  {Delacr\'etaz}}, \bibinfo {author} {\bibfnamefont {B.}~\bibnamefont
  {Gout\'eraux}}, \bibinfo {author} {\bibfnamefont {S.~A.}\ \bibnamefont
  {Hartnoll}},\ and\ \bibinfo {author} {\bibfnamefont {A.}~\bibnamefont
  {Karlsson}},\ }\href {https://doi.org/10.1103/PhysRevB.96.195128} {\bibfield
  {journal} {\bibinfo  {journal} {Phys. Rev. B}\ }\textbf {\bibinfo {volume}
  {96}},\ \bibinfo {pages} {195128} (\bibinfo {year} {2017})}\BibitemShut
  {NoStop}%
\bibitem [{\citenamefont {Andrade}\ \emph {et~al.}(2018)\citenamefont
  {Andrade}, \citenamefont {Krikun}, \citenamefont {Schalm},\ and\
  \citenamefont {Zaanen}}]{andrade2018}%
  \BibitemOpen
  \bibfield  {author} {\bibinfo {author} {\bibfnamefont {T.}~\bibnamefont
  {Andrade}}, \bibinfo {author} {\bibfnamefont {A.}~\bibnamefont {Krikun}},
  \bibinfo {author} {\bibfnamefont {K.}~\bibnamefont {Schalm}},\ and\ \bibinfo
  {author} {\bibfnamefont {J.}~\bibnamefont {Zaanen}},\ }\href
  {https://doi.org/10.1038/s41567-018-0217-6} {\bibfield  {journal} {\bibinfo
  {journal} {Nature Physics}\ }\textbf {\bibinfo {volume} {14}},\ \bibinfo
  {pages} {1049–1055} (\bibinfo {year} {2018})}\BibitemShut {NoStop}%
\bibitem [{\citenamefont {Caprara}\ \emph {et~al.}(2022)\citenamefont
  {Caprara}, \citenamefont {Castro}, \citenamefont {Mirarchi}, \citenamefont
  {Seibold},\ and\ \citenamefont {Grilli}}]{caprara2022}%
  \BibitemOpen
  \bibfield  {author} {\bibinfo {author} {\bibfnamefont {S.}~\bibnamefont
  {Caprara}}, \bibinfo {author} {\bibfnamefont {C.~D.}\ \bibnamefont {Castro}},
  \bibinfo {author} {\bibfnamefont {G.}~\bibnamefont {Mirarchi}}, \bibinfo
  {author} {\bibfnamefont {G.}~\bibnamefont {Seibold}},\ and\ \bibinfo {author}
  {\bibfnamefont {M.}~\bibnamefont {Grilli}},\ }\href
  {https://doi.org/10.1038/s42005-021-00786-y} {\bibfield  {journal} {\bibinfo
  {journal} {Communications Physics}\ }\textbf {\bibinfo {volume} {5}},\
  \bibinfo {pages} {10} (\bibinfo {year} {2022})}\BibitemShut {NoStop}%
\bibitem [{\citenamefont {Fujita}\ \emph {et~al.}(2004)\citenamefont {Fujita},
  \citenamefont {Goka}, \citenamefont {Yamada}, \citenamefont {Tranquada},\
  and\ \citenamefont {Regnault}}]{fujita2004}%
  \BibitemOpen
  \bibfield  {author} {\bibinfo {author} {\bibfnamefont {M.}~\bibnamefont
  {Fujita}}, \bibinfo {author} {\bibfnamefont {H.}~\bibnamefont {Goka}},
  \bibinfo {author} {\bibfnamefont {K.}~\bibnamefont {Yamada}}, \bibinfo
  {author} {\bibfnamefont {J.~M.}\ \bibnamefont {Tranquada}},\ and\ \bibinfo
  {author} {\bibfnamefont {L.~P.}\ \bibnamefont {Regnault}},\ }\href
  {https://doi.org/10.1103/PhysRevB.70.104517} {\bibfield  {journal} {\bibinfo
  {journal} {Phys. Rev. B}\ }\textbf {\bibinfo {volume} {70}},\ \bibinfo
  {pages} {104517} (\bibinfo {year} {2004})}\BibitemShut {NoStop}%
\end{thebibliography}
\providecommand{\noopsort}[1]{}\providecommand{\singleletter}[1]{#1}%

\onecolumngrid
\newpage
\appendix*
\section{Supplementary Materials}
\renewcommand{\thefigure}{S\arabic{figure}}
\setcounter{figure}{0}

\subsection{Methods}
\subsubsection{Hubbard model}
The Hubbard model Hamiltonian is
\begin{equation}
    H = -\sum_{ij\sigma} t_{ij} c_{i\sigma}^\dagger c_{j\sigma} + U \sum_{i}\hat{n}_{i\uparrow}\hat{n}_{i\downarrow} - \mu\sum_{i\sigma}\hat{n}_{i\sigma}, \label{eq:hub_model}
\end{equation}
where $t_{ij}$ are electron hopping matrix elements which parametrize the kinetic energy (here, we assume only non-zero nearest- and next-nearest-neighbor hopping matrix elements denoted as $t$ and $t'$); $U$ is the on-site Coulomb repulsion; $\mu$ is the chemical potential controlling the number of electrons; $c^{\dagger}_{i\sigma}$ ($c_{i\sigma}$) are electron creation (annihilation) operators for electrons at site $i$ with spin $\sigma$; and $\hat{n}_{i\sigma} = c^\dagger_{i\sigma}c_{i\sigma}$ is the number operator. Simulation parameters for the results presented in this paper are summarized in Table \ref{table:1}. 

\subsubsection{Determinant quantum Monte Carlo}
To study the Hubbard model, we perform determinant quantum Monte Carlo (DQMC) simulations, which is a numerically exact method capable of calculating finite temperature and dynamical properties ~\cite{white1989,bss1981}. The DQMC algorithm is limited by the fermion sign problem, rendering the difficulty of the calculations exponentially growing with lattice size, interaction strength and inverse temperature. Hence the simulation is restricted to elevated temperature regimes.

The chemical potential is tuned to achieve the desired doping levels with an accuracy of $\sim 10^{-4}$. $50000$ warm-up sweeps are performed for the Monte Carlo simulation. Due to the exponential decay of charge correlations and the extremely small fermion sign, which can be as low as $\sim 0.015$, a large amount of data is needed to show multiple stripe domains with good statistics. We use up to $\sim 10000$ independently seeded Markov chains and up to $10$ million measurement sweeps, amounting to a total number of measurements between $10$ to $100$ billion, to resolve charge stripe modulations with magnitude as small as $10^{-5}$ with standard errors on the order of $10^{-6}$. Simulations details for each parameter set may be found in Table~\ref{table:1}. 

\begin{table}[h]
\begin{tabular}{|l|l|l|l|l|l|l|l|l|}
\hline
\multicolumn{1}{|c|}{Cluster size} & \multicolumn{1}{c|}{$U/t$} & \multicolumn{1}{c|}{$t'/t$} & \multicolumn{1}{c|}{Doping} & Temperature    & Bins          & \begin{tabular}[c]{@{}l@{}}Equal-time\\ measurements/bin\end{tabular} & \begin{tabular}[c]{@{}l@{}}Unequal-time \\ measurements/bin\end{tabular} & Figures                \\ \hline
$12\times4$                        & $6$ & $-0.25$                     & $0.125$                     & $0.22$         & $7984$        & $4.5\times10^7$                                                       & $2.5\times10^6$                                                          & 1a, S1, S3            \\ \hline
$8\times8$                         & $6$ & $-0.25$                     & $0.125$                     & $0.22$         & $9571$        & $1.6\times10^7$                                                       & $2\times10^6$                                                            & 1b                    \\ \hline
$12\times4$                        & $6$ & $-0.25$                     & $[0.05, 0.2]$               & $0.22$         & Up to $2867$  & $5\times10^7$                                                         & $5\times10^6$                                                            & 2, S5                 \\ \hline
$16\times4$                        & $6$ & $-0.25$                     & $0.125$                     & $0.22$         & $1946$        & $1.6\times10^7$                                                       & $0$                                                                      & 3                     \\ \hline
$16\times4$                        & $6$ & $-0.25$                     & $[0.042, 0.25]$             & $[0.22, 0.67]$ & Up to $9997$ & Up to $7\times10^6$                                                   & $2.5\times10^5$                                                          & 4, S2, S3, S4, S6, S7 \\ \hline
$8\times4, 10\times4$              & $6$ & $-0.25$                     & $0.125$                     & $0.22$         & $2000$        & $4.5\times10^7$                                                       & $2.5\times10^6$                                                          & S3                    \\ \hline
\end{tabular}

\caption{Simulation parameters for the results presented in this paper.}
\label{table:1}
\end{table}

\subsubsection{Error analysis}
For DQMC simulations, we use jackknife resampling to estimate the standard errors. Typical errors in the susceptibilities are $\sim 10^{-6}$. The imaginary time discretization in DQMC algorithm introduces a systematic Trotter error. We keep the time slice $\Delta\tau \le 0.125/t$ so that the Trotter error is negligible, as confirmed by previous DQMC studies~\cite{huang2017}.

\subsubsection{Spin-spin-hole correlation function}
Evaluating multi-point correlation functions in DQMC is accomplished by applying Wick's theorem each time a measurement is performed~\cite{bss1981,white1989}. The three-point spin-spin-hole correlation function plotted in Fig.~\ref{fig:3} involves terms with eight fermion operators. To enumerate the Wick contractions of these terms, we utilize a symbolic algebra system to automatically generate measurement code for this observable. For reference, the Wick decomposition is
\begin{equation}
\begin{split}
\ev{m^z_{i} m^z_{j} h_{k}}_s = \frac{1}{4} [
(
&(1 - g_{i i}^{\uparrow}) (1 - g_{j j}^{\uparrow}) g_{k k}^{\uparrow} g_{k k}^{\downarrow} -(1 - g_{i i}^{\uparrow}) (\delta_{j k} - g_{k j}^{\uparrow}) g_{j k}^{\uparrow} g_{k k}^{\downarrow} \\
&+(\delta_{i j} - g_{j i}^{\uparrow}) g_{i j}^{\uparrow} g_{k k}^{\uparrow} g_{k k}^{\downarrow} +(\delta_{i j} - g_{j i}^{\uparrow}) g_{i k}^{\uparrow} (\delta_{j k} - g_{k j}^{\uparrow}) g_{k k}^{\downarrow} \\
&-(\delta_{i k} - g_{k i}^{\uparrow}) g_{i j}^{\uparrow} g_{j k}^{\uparrow} g_{k k}^{\downarrow} -(\delta_{i k} - g_{k i}^{\uparrow}) g_{i k}^{\uparrow} (1 - g_{j j}^{\uparrow}) g_{k k}^{\downarrow}
)\\
- (
&(1 - g_{i i}^{\uparrow}) (1 - g_{j j}^{\downarrow}) g_{k k}^{\uparrow} g_{k k}^{\downarrow} -(1 - g_{i i}^{\uparrow}) (\delta_{j k} - g_{k j}^{\downarrow}) g_{j k}^{\downarrow} g_{k k}^{\uparrow} \\
&-(\delta_{i k} - g_{k i}^{\uparrow}) g_{i k}^{\uparrow} (1 - g_{j j}^{\downarrow}) g_{k k}^{\downarrow} +(\delta_{i k} - g_{k i}^{\uparrow}) g_{i k}^{\uparrow} (\delta_{j k} - g_{k j}^{\downarrow}) g_{j k}^{\downarrow}
)\\
- (
&(1 - g_{i i}^{\downarrow}) (1 - g_{j j}^{\uparrow}) g_{k k}^{\uparrow} g_{k k}^{\downarrow} -(1 - g_{i i}^{\downarrow}) (\delta_{j k} - g_{k j}^{\uparrow}) g_{j k}^{\uparrow} g_{k k}^{\downarrow} \\
&-(\delta_{i k} - g_{k i}^{\downarrow}) g_{i k}^{\downarrow} (1 - g_{j j}^{\uparrow}) g_{k k}^{\uparrow} +(\delta_{i k} - g_{k i}^{\downarrow}) g_{i k}^{\downarrow} (\delta_{j k} - g_{k j}^{\uparrow}) g_{j k}^{\uparrow}
)\\
+ (
&(1 - g_{i i}^{\downarrow}) (1 - g_{j j}^{\downarrow}) g_{k k}^{\uparrow} g_{k k}^{\downarrow} -(1 - g_{i i}^{\downarrow}) (\delta_{j k} - g_{k j}^{\downarrow}) g_{j k}^{\downarrow} g_{k k}^{\uparrow} \\
&+(\delta_{i j} - g_{j i}^{\downarrow}) g_{i j}^{\downarrow} g_{k k}^{\uparrow} g_{k k}^{\downarrow} +(\delta_{i j} - g_{j i}^{\downarrow}) g_{i k}^{\downarrow} (\delta_{j k} - g_{k j}^{\downarrow}) g_{k k}^{\uparrow} \\
&-(\delta_{i k} - g_{k i}^{\downarrow}) g_{i j}^{\downarrow} g_{j k}^{\downarrow} g_{k k}^{\uparrow} -(\delta_{i k} - g_{k i}^{\downarrow}) g_{i k}^{\downarrow} (1 - g_{j j}^{\downarrow}) g_{k k}^{\uparrow}
) ].
\end{split}
\end{equation}
Here, $g_{a b}^\sigma = \ev{c_{a\sigma} c_{b\sigma}^\dagger}_s$ is the single particle Green's function measured for the auxiliary field configuration $s$.

\subsubsection{Stripe strength}\label{sec:stripe_strength}
\begin{figure*}
    \includegraphics[scale=0.5]{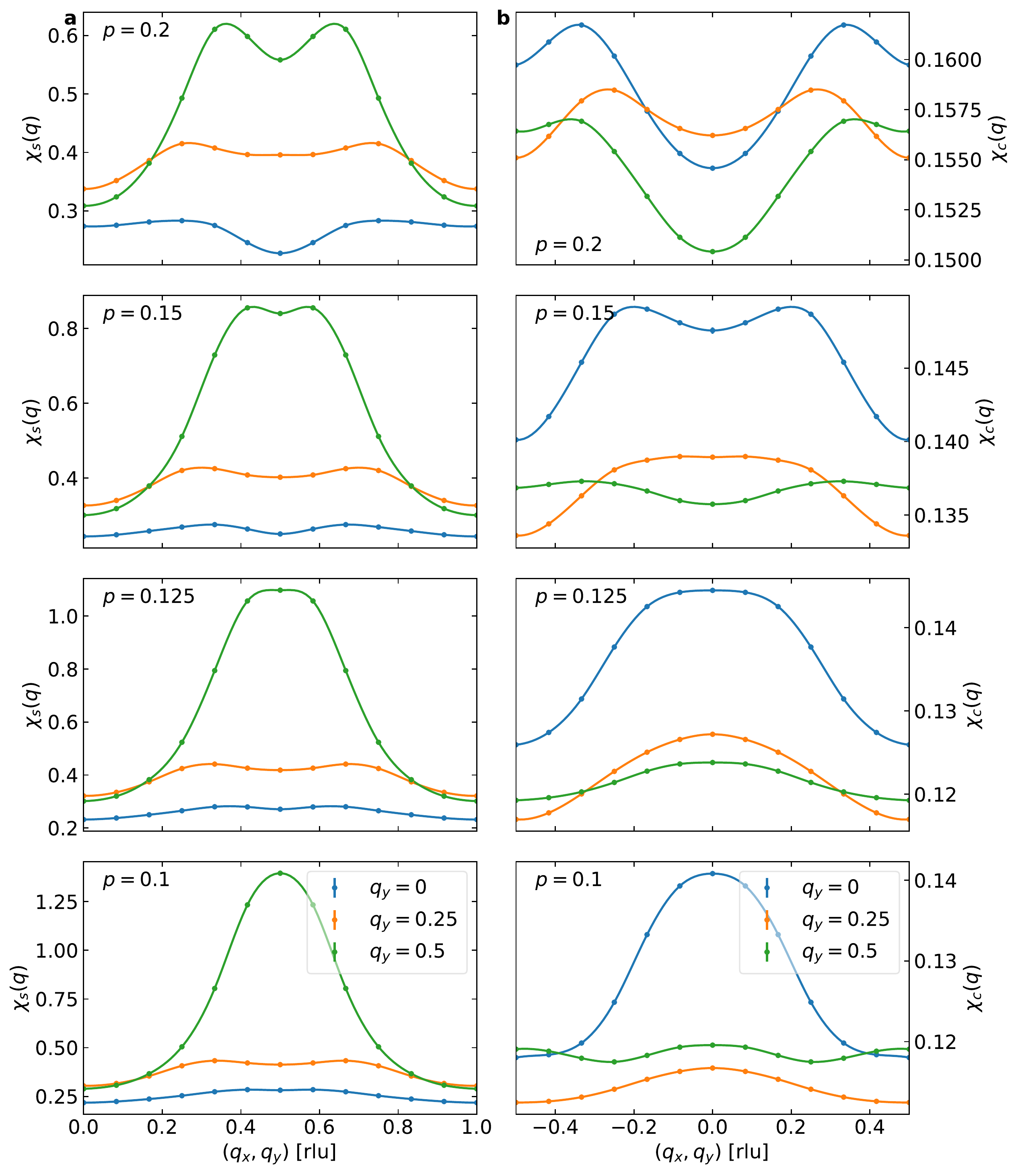}
    \caption{Momentum-space \textbf{a,} spin and \textbf{b,} charge susceptibilities for different hole dopings, plotted along the $q_x$ direction for $q_y=0, 0.25, 0.5$ (rlu).}
    \label{fig:nnzzqw0s}
\end{figure*}

To characterize the strength of fluctuating stripes, we consider in Fig.~\ref{fig:phase_diagram} the value of the spin and charge susceptibilities at the nearest vertical neighbor, $\chi_s(\mathbf{r} = \hat{y})$ and $\chi_c(\mathbf{r} = \hat{y})$, as estimates of the magnitude of the fluctuating stripes. Although we choose this metric for its simplicity, an advantage is that on rectangular clusters with width $4$, 
\begin{equation}
\chi(\mathbf{r} = \hat{y}) \propto \sum_{q_x} \chi(\mathbf{q}=(q_x, 0)) - \chi(\mathbf{q}=(q_x, 0.5)). \label{eq:chiry_FT_}
\end{equation}
We find empirically that $\chi_s(\mathbf{q}=(q_x, 0))$ and $\chi_c(\mathbf{q}=(q_x, 0.5))$ are relatively independent of $q_x$ [Fig. \ref{fig:nnzzqw0s}]. Therefore, the right hand side of \eqref{eq:chiry_FT_} corresponds approximately to the weight under the peaks seen in Fig.~\ref{fig:2}{\bf d,e} with a constant background subtracted. For doping concentrations or temperatures where $\chi_c(\mathbf{r} = \hat{y}) < 0$, the patterns in the charge susceptibility do not resemble stripes [\ref{fig:Tdep}].

\subsection{Finite size effects}
\begin{figure*}
    \includegraphics[scale=0.4]{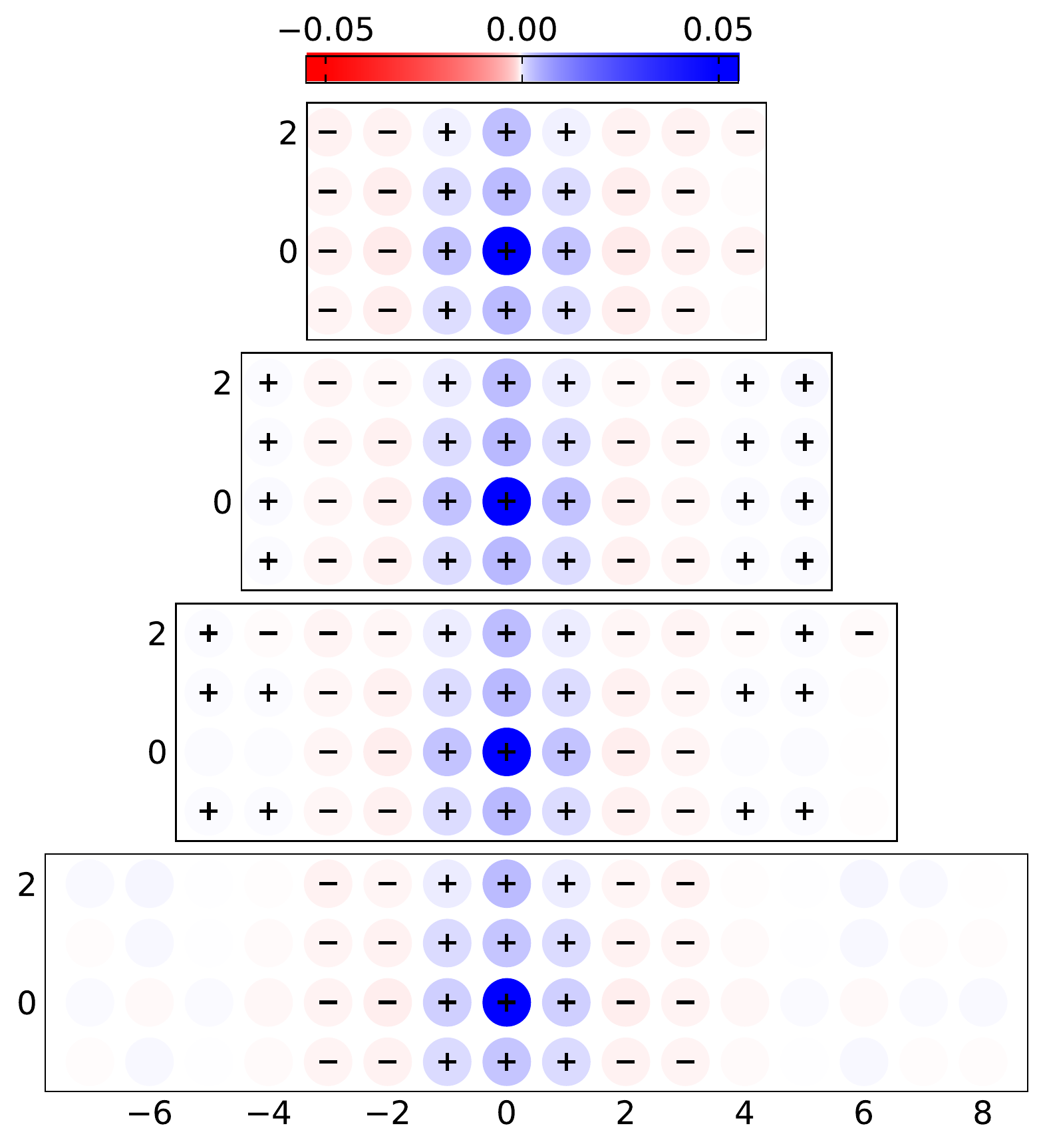}
    \caption{Charge susceptibilities for different horizontal lengths of the rectangular cluster, $N_x = 8, 10, 12, 16$, with parameters $U/t=6$, $t'/t=-0.25$, $p = 0.125$ and $T/t=0.22$. Locations for the sign change for charge stripes are the same across different cluster lengths, indicating little effect due to the finite cluster size. The worsened fermion sign problem for larger cluster ($N_x=16$) prevents a clear resolution of the additional positive correlation domains of charge stripes, which are present in smaller clusters.}
    \label{fig:Nx_dep}
\end{figure*}

To ensure that our results are not artifacts of the finite cluster size, we run the simulation for a range of lengths of the rectangular clusters. Fig. \ref{fig:Nx_dep} shows the real-space charge susceptibilities for lengths of $N_x = 8, 10, 12, 16$ and $N_y=4$. The stripe modulation is clearly present in all four types of clusters, with similar periodicity as indicated by the positions of the lines of negative correlations. The limited cluster size does not appear to have severe effect on observing signatures of charge stripes, possibly due to the short correlation lengths. The details of the correlation functions may be slightly sensitive to the lattice geometries. For example, at a hole doping of $1/8$ with a charge stripe periodicity of $\sim5$, the lattice with length $10$ commensurate with the periodicity shows perfect domains of positive and negative correlations running vertically across the lattice, while in lattices with other lengths such as $N_x=12$ there are negative signs in predominantly positive domains. The fermion sign is reduced with increasing system size: $\langle sign \rangle \approx 0.13, 0.08, 0.06, 0.02$ for $N_x = 8, 10, 12, 16$. Due to the worsened sign problem as cluster length increases, it becomes more difficult to resolve the second positive domain with reasonable standard error. For $16\times4$, the standard error is too big for any positive correlations beyond the first stripe domain to be unambiguously resolved.

\subsection{Susceptibility fitting}\label{sec:fitting}
\begin{figure*}
    \includegraphics[scale=0.4]{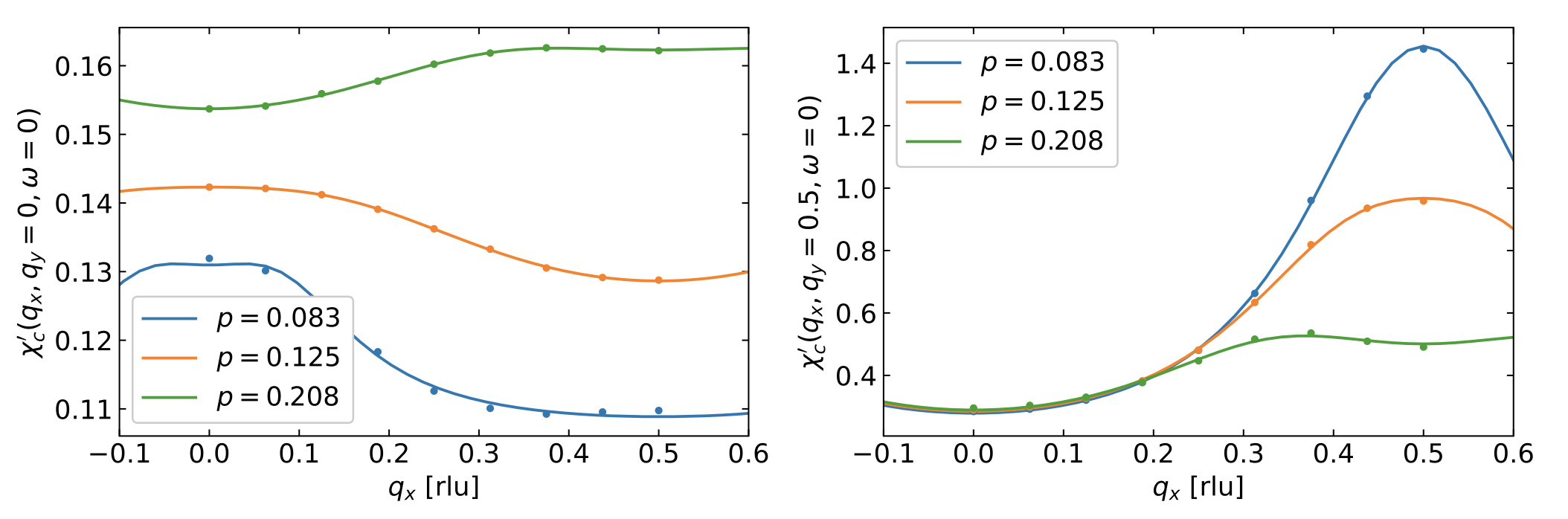}
    \caption{Example fits to the susceptibilities in momentum space, using double Lorentzian curves centered around $(\pi, \pi)$ and $(0, 0)$ for spin and charge respectively. The Lorentzian function is repeated over several Brillouin zones.}
    \label{fig:fits}
\end{figure*}

The momentum-space susceptibilities show incommensurate peaks that split from commensurate wave vectors $(\pi,\pi)$ and $(0,0)$ for spin and charge stripes, respectively. We perform fits to the susceptibilities for a quantitative estimate of the stripe incommensurability. For the fitting function we choose periodic (repeated over several Brillouin zones) double Lorentzian functions.
\begin{equation}
    \chi_{fit}(q) = A \sum_G \frac{1}{(q - Q_{DW} + G)^2 + \Gamma^2} + \frac{1}{(q + Q_{DW} + G)^2 + \Gamma^2}.
\end{equation}
Here, $Q_{DW}$ is the charge or spin stripe wavevector, $\Gamma$ is the peak width, and $A$ controls the height. $G$ shifts $q$ to different Brillouin zones. In our fits we include contributions from 20 values of $G$. The periodicity of $\chi_{fit}(q)$ allows for successful fits to the entirety of the data, without arbitrary choices of points to exclude. An additional constant is added as a free parameter for fits to the charge susceptibilities to account for a momentum-independent background. As shown in Fig.\ref{fig:2} and \ref{fig:fits}, the susceptibilities are well fit for a range of dopings.

\subsection{Temperature Dependence}
\begin{figure*}
    \includegraphics[scale=0.4]{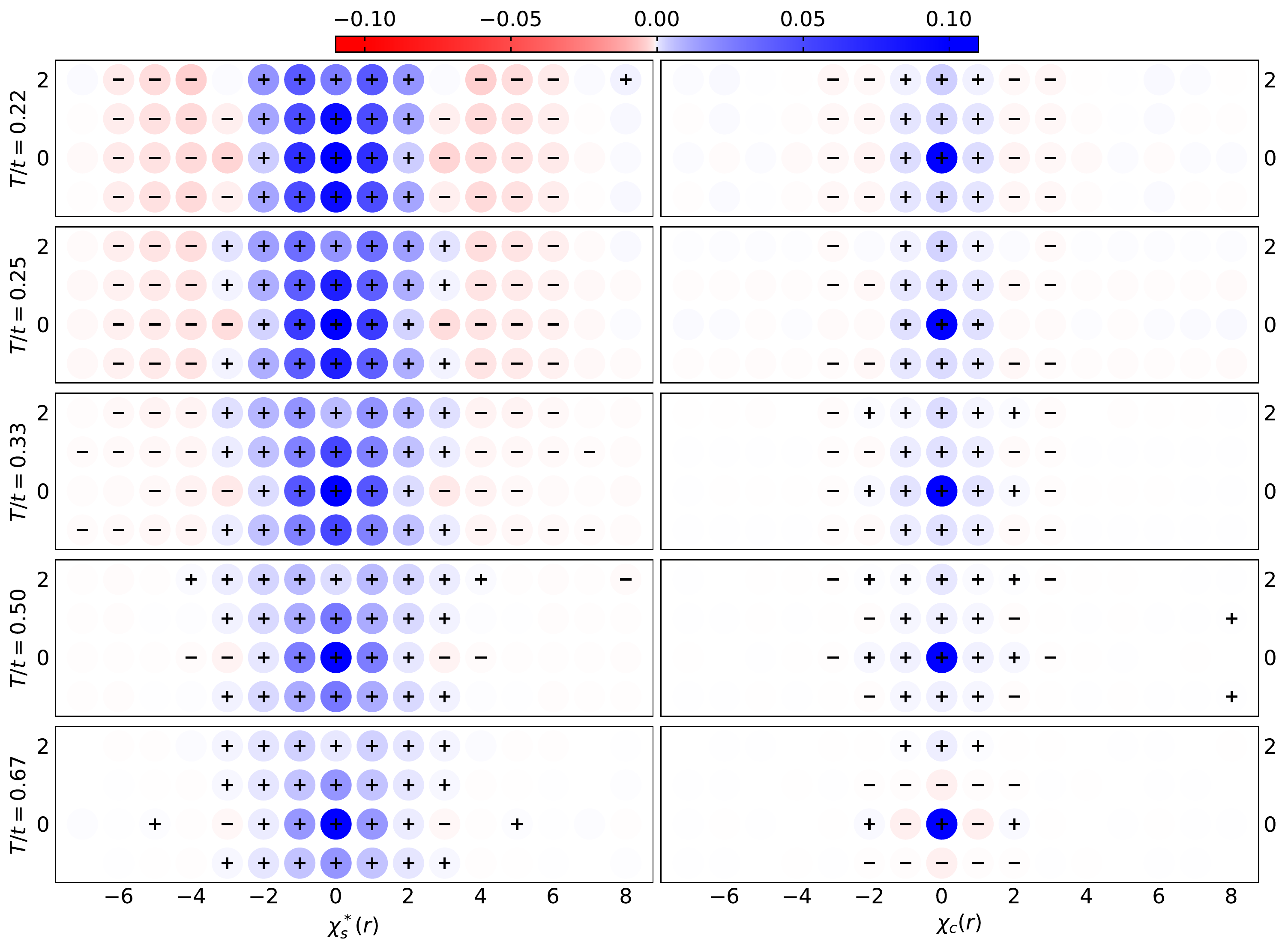}
    \caption{Spin and charge correlation functions at zero frequency for various temperatures with parameters $U/t=6$, $t'/t=-0.25$, $p = 0.125$.}
    \label{fig:Tdep}
\end{figure*}

The temperature dependence of stripes are summarized in figure \ref{fig:Tdep}. While increasing temperature leads to larger fermion signs, it also results in decreased correlation lengths in both spin and charge susceptibilities, as evidenced through the decreased values of the susceptibilities at large distances. We observe that, consistent with neutron scattering experiments on La\textsubscript{1.875}Ba\textsubscript{0.125}CuO\textsubscript{4} ~\cite{fujita2004}, the periodicity of spin stripes increases with temperature. This could indicate that at higher temperatures a longer cluster is necessary to resolve any potential stripe correlations. We generally find incommensurate spin correlations below roughly $T/t\approx0.6$; at higher temperatures no clear stripe signatures are visible due to the reduced magnetic correlation and the longer periodicity of potential stripe modulation. For charge, the negative correlation regions are present up to temperature between $T/t = 0.33$ and $0.5$. With increasing temperature the stripes appear more fluctuating, indicated by the leakage of negative correlations into a positive domain or vice versa. At and above $T/t=0.67$, the pattern of the charge susceptibility does not resemble stripes.

\subsection{Pseudogap crossover temperature $T^*$}\label{sec:PG}
\begin{figure*}
    \includegraphics[scale=0.7]{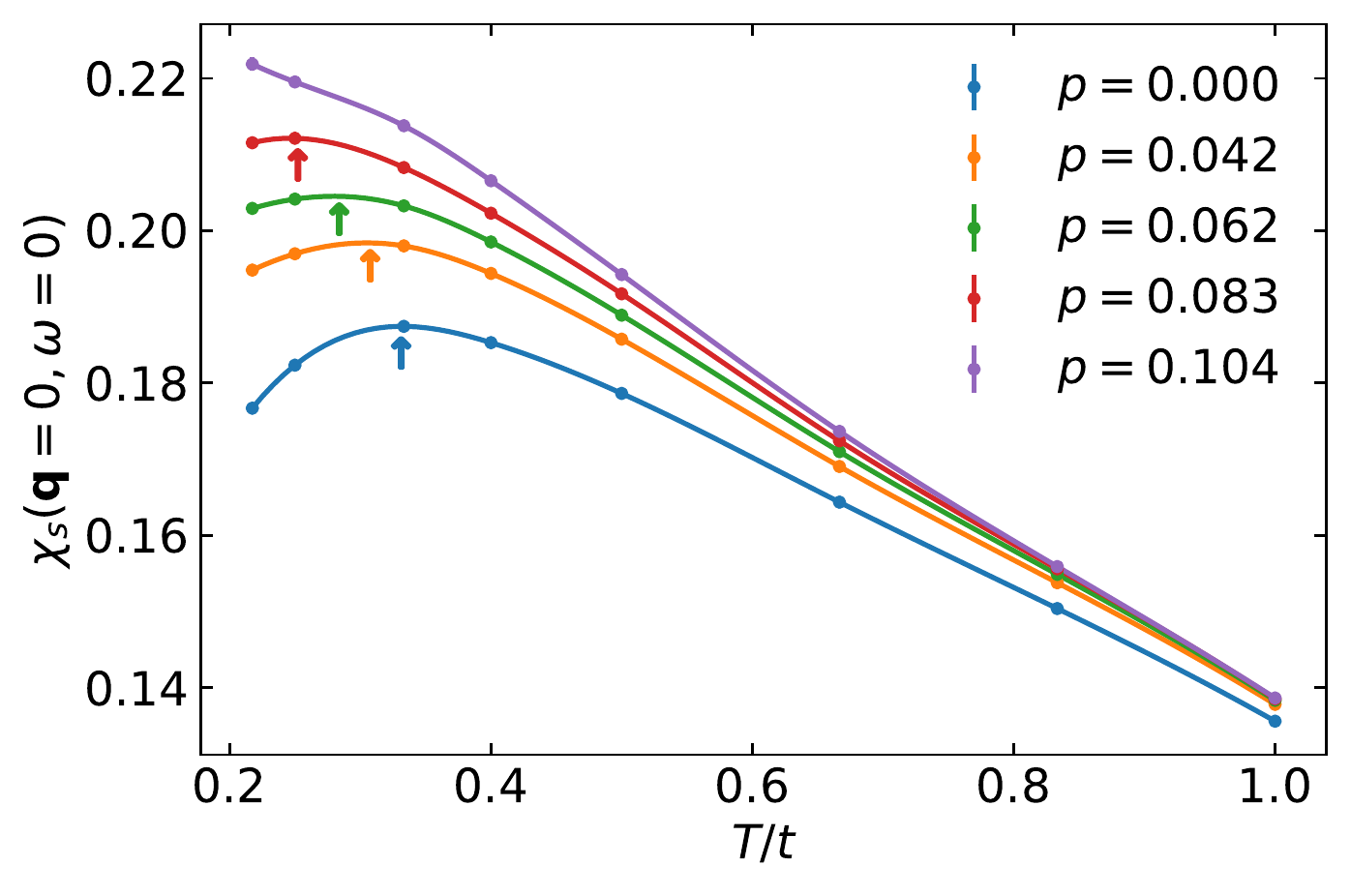}
    \caption{Spin (Pauli) susceptibility, or Knight shift, as a function of temperature for different dopings with parameters $U/t=6$, $t'/t=-0.25$. Solid lines are cubic spline fits to the data. Arrows indicate peak positions. The peak cannot be tracked for dopings at and larger than $p=0.104$ due to the fermion sign problem.}
    \label{fig:pauli}
\end{figure*}

One of the indications for a crossover into the pseudogap regime is a peak in the Knight shift from nuclear magnetic resonance experiment. In our calculations, the Knight shift is given by the Pauli spin susceptibility
\begin{equation}
    \chi_s(\mathbf{q}=0, \omega=0) = \sum_{\mathbf{r}}\int_0^\beta \dd{\tau} \ev{m^z_{\mathbf{r}}(\tau) m^z_0}.
\end{equation}
The Knight shifts for different dopings as a function of temperature are shown in Fig.~\ref{fig:pauli}. The pseudogap crossover temperature $T^*$ is estimated by performing cubic spline fit to the data. Our calculations found peaks in the Knight shift at low temperatures that decrease with doping. At and beyond $\sim10\%$ doping, the peak temperature is below $T/t=0.2$.

\newpage

\begin{figure*}
    \includegraphics[scale=0.4]{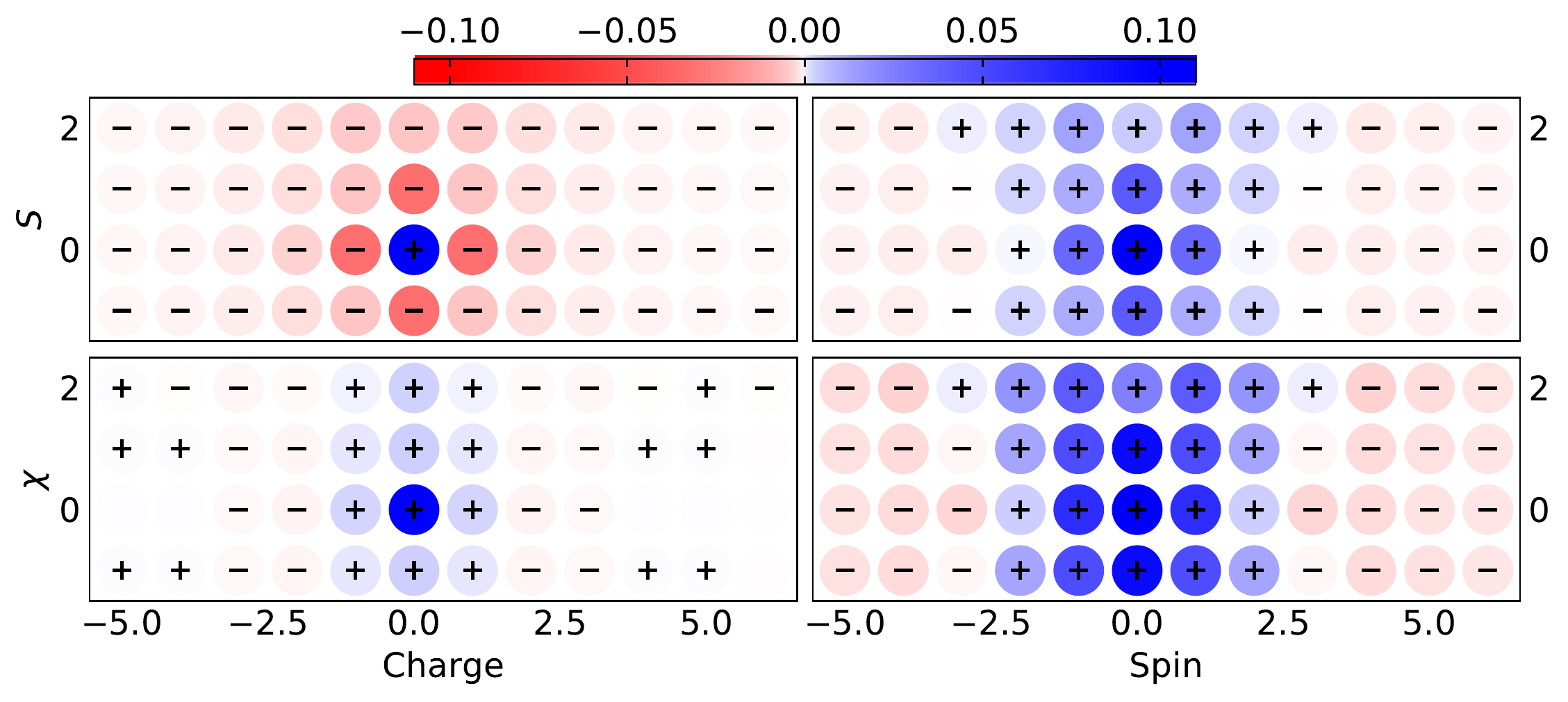}
    \caption{Equal-time correlation functions ($S$, upper column) and zero-frequency susceptibilities ($\chi$, lower column) for spin and charge on a $12\times4$ cluster.}
    \label{fig:chi_S_12x4}
\end{figure*}

\begin{figure*}
    \includegraphics[scale=0.4]{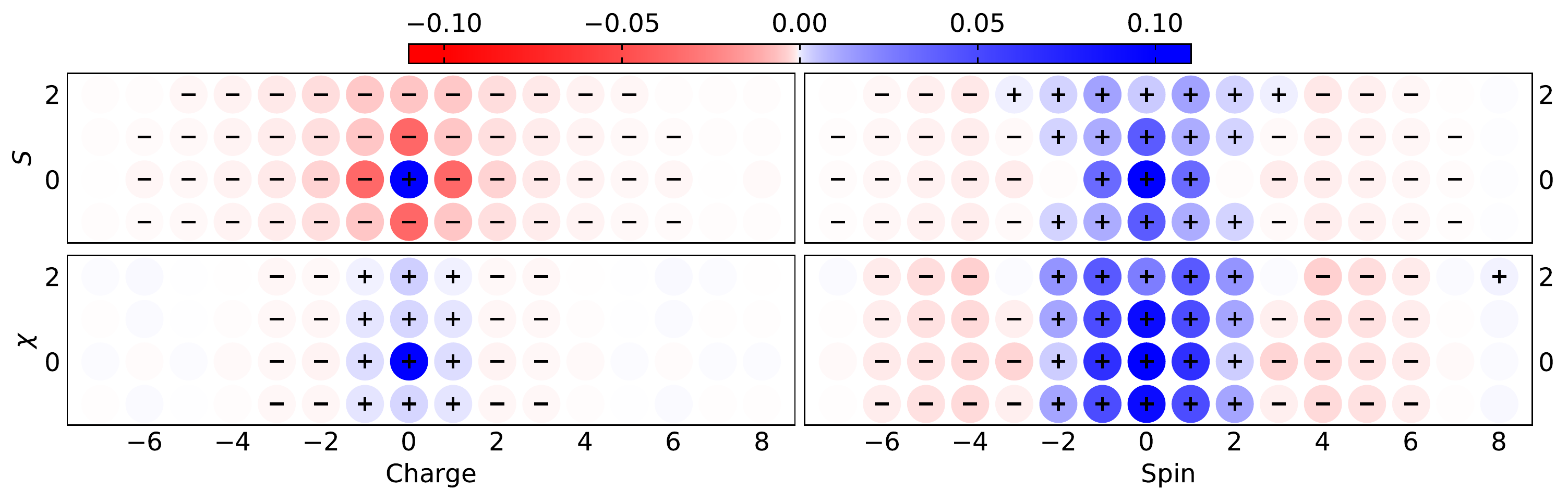}
    \caption{Equal-time correlation functions ($S$, upper column) and zero-frequency susceptibilities ($\chi$, lower column) for spin and charge on a $16\times4$ cluster.}
    \label{fig:chi_S_16x4}
\end{figure*}

\end{document}